\documentclass[paper]{JHEP3}
\pdfoutput=1
\usepackage{amsmath,amssymb,amsthm,amscd,graphicx}
\usepackage{multirow}

\theoremstyle{definition}

\newcommand{\CC}{{\cal C}}

\newcommand{\CF}{{\cal F}}

\newcommand{\CH}{{\cal H}}

\newcommand{\CK}{{\cal K}}
\newcommand{\CL}{{\cal L}}

\newcommand{\CO}{{\cal O}}
\newcommand{\CP}{{\cal P}}

\newcommand{\CR}{{\cal R}}
\newcommand{\CS}{{\cal S}}

\newcommand{\CU}{{\cal U}}
\newcommand{\CV}{{\cal V}}
\newcommand{\CW}{{\cal W}}

\def\IZ{{\mathbb Z}}
\def\IR{{\mathbb R}}
\def\IC{{\mathbb C}}
\def\IP{{\mathbb P}}
\def\IT{{\mathbb T}}
\def\IS{{\mathbb S}}

\newcommand{\tr}{{\rm Tr}}
\newcommand{\re}{{\rm e}}
\newcommand{\ri}{{\rm i}}
\newcommand{\rd}{{\rm d}}

\newcommand{\bra}{\left\langle}
\newcommand{\ket}{\right\rangle}

\newcommand{\be}{\begin{equation}}
\newcommand{\ee}{\end{equation}}
\newcommand{\ba}{\begin{aligned}}
\newcommand{\ea}{\end{aligned}}
\newcommand{\ben}{\begin{eqnarray}\displaystyle}
\newcommand{\een}{\end{eqnarray}}
\newcommand{\nn}{\nonumber}

\newcommand{\sectiono}[1]{\section{#1}\setcounter{equation}{0}}

\newdimen\tableauside\tableauside=1.0ex
\newdimen\tableaurule\tableaurule=0.4pt
\newdimen\tableaustep
\def\phantomhrule#1{\hbox{\vbox to0pt{\hrule height\tableaurule width#1\vss}}}
\def\phantomvrule#1{\vbox{\hbox to0pt{\vrule width\tableaurule height#1\hss}}}
\def\sqr{\vbox{%
  \phantomhrule\tableaustep
  \hbox{\phantomvrule\tableaustep\kern\tableaustep\phantomvrule\tableaustep}%
  \hbox{\vbox{\phantomhrule\tableauside}\kern-\tableaurule}}}
\def\squares#1{\hbox{\count0=#1\noindent\loop\sqr
  \advance\count0 by-1 \ifnum\count0>0\repeat}}
\def\tableau#1{\vcenter{\offinterlineskip
  \tableaustep=\tableauside\advance\tableaustep by-\tableaurule
  \kern\normallineskip\hbox
    {\kern\normallineskip\vbox
      {\gettableau#1 0 }%
     \kern\normallineskip\kern\tableaurule}%
  \kern\normallineskip\kern\tableaurule}}
\def\gettableau#1{\ifnum#1=0\let\next=\null\else
\squares{#1}\let\next=\gettableau\fi\next}

\def\IE{\mathbb{E}}

\newcommand{\figref}[1]{Fig.~\protect\ref{#1}}
\title{Torus knots and mirror symmetry}

\author{Andrea Brini$^{a}$, Bertrand Eynard$^{b,c}$ and Marcos Mari\~no$^{a}$
\\
$^a$D\'epartement de Physique Th\'eorique et Section de Math\'ematiques,\\
Universit\'e de Gen\`eve, Gen\`eve, CH-1211 Switzerland\\
\\
$^b$Department of Theoretical Physics, CERN\\
Gen\`eve, CH-1213 Switzerland\\
\\
$^c$ Service de Physique Th\'eorique de Saclay\\
F-91191 Gif-sur-Yvette Cedex, France
}

\preprint{CERN-2011/103 \\
IPHT-T11/134}
\abstract{We propose a spectral curve describing torus knots and links in the B-model. In particular, 
the application of the topological recursion to this curve generates all their colored HOMFLY invariants. The curve is obtained by exploiting the full 
${\rm Sl}(2,\IZ)$ symmetry of the spectral curve of the resolved conifold, and should be regarded as the mirror of the topological D-brane associated to torus knots 
in the large $N$ Gopakumar--Vafa duality. Moreover, we derive the curve as the large $N$ limit of the matrix model computing torus knot invariants.}    

\begin{document}
\tableofcontents

\sectiono{Introduction}

One of the most surprising 
consequences of the Gopakumar--Vafa duality \cite{gv} is that Chern--Simons invariants of knots and links in the three-sphere can be described by A-model 
open topological strings on the resolved conifold \cite{ov} (see \cite{mmten} for a recent review). The boundary conditions for the open strings are set by 
a Lagrangian submanifold associated to the knot or link. By mirror symmetry, an equivalent description should exist in terms of open strings in the B-model, 
where the boundary conditions are set by holomorphic submanifolds. 

This conjectural equivalence between knot theory and Gromov--Witten theory has
been implemented and tested in detail for the (framed) unknot and the Hopf
link. For the framed unknot there is a candidate Lagrangian submanifold in the
A-model \cite{ov}. Open Gromov--Witten invariants for this geometry can be
defined and calculated explicitly by using for example the theory of the topological vertex \cite{akmv}, and they agree with the corresponding Chern--Simons invariants (see for example \cite{zhou} for a recent study and references to earlier work). The framed unknot can be also studied in the B-model \cite{akv,mv}. As usual in local mirror symmetry, the mirror is an algebraic curve in $\IC^* \times \IC^*$, and the invariants of the framed unknot can be computed as open topological string amplitudes in this geometry using the formalism of \cite{mmopen,bkmp}. The Hopf link can be also understood in the framework of topological strings and Gromov--Witten theory (see for example \cite{ikp}).  

In spite of all these results, there has been little progress in extending the conjectural equivalence between knot theory and string theory to other knots and links. There have been 
important indirect tests based on integrality properties (see \cite{mmten} for a review), but no concrete string theory calculation of Chern--Simons invariants of knots and links 
has been proposed beyond the unknot and the Hopf link, 
even for the trefoil (which is the simplest non-trivial knot).   

In this paper we make a step to remedy this situation, and we provide a computable, B-model description of all torus knots and links. Torus knots and links are very special and simple, but they are an important testing ground in knot theory and Chern--Simons theory. As we will see, our B-model description does not involve radically new ingredients, but it definitely extends the string/knot dictionary beyond the simple examples known so far. 

Our proposal is a simple and natural generalization of \cite{akv}. It is known
that for B-model geometries that describe mirrors of local Calabi--Yau
threefolds, and are thus described by a mirror Riemann surface, there is an
${\rm Sl}(2,\IZ)$
action that rotates the B-model open string moduli with the reduction of the
holomorphic three-form on the spectral curve; this action is a symmetry of the
closed string sector. For open strings, it was proposed in \cite{akv} that the unknot with $f$ units of framing is obtained by acting with the ${\rm Sl}(2,\IZ)$ transformation $T^f$ on the spectral curve of the resolved conifold (here $T$ denotes the standard generator of the modular group), but no interpretation 
was given for a more general modular transformation. As we will show in this
paper, the B-model geometry corresponding to a $(Q,P)$ torus knot is simply
given by a {\it general}  ${\rm Sl}(2,\IZ)$ transformation of the spectral curve describing the resolved conifold. This proposal clarifies 
the meaning of general symplectic transformations of spectral curves, which play a crucial r\^ole in the formalism of \cite{eo}. Moreover, it is in perfect agreement with the 
Chern--Simons realization of the Verlinde algebra. In this realization, one shows \cite{llr} that torus knots are related to the (framed) unknot by a general 
symplectic transformation. Our result can be simply stated by saying that the natural ${\rm Sl}(2,\IZ)$ action on torus knots in the canonical quantization of Chern--Simons theory is equivalent to the ${\rm Sl}(2,\IZ)$ reparametrization of the spectral curve. 

In practical terms, the above procedure associates a spectral curve to each torus knot or link. Their colored $U(N)$ invariants can then be computed 
systematically by applying the topological recursion of \cite{eo} to the
spectral curve, exactly as in \cite{bkmp}. In this description, the $(P,Q)$
torus knot comes naturally equipped with a fixed framing of $QP$ units, just
as in Chern--Simons theory \cite{llr}. As a spinoff of this study, we obtain a
formula for the HOMFLY polynomial of a $(Q,P)$ torus knot in terms of
$q$-hypergeometric polynomials and recover the results of \cite{gorsky}. 
 
Our result for the torus knot spectral curve is very natural, but on top of
that we can actually derive it. This is because the colored $U(N)$ invariants
of torus knots admit a matrix integral representation, as first pointed out in
the $SU(2)$ case in \cite{lr}. The calculation of \cite{lr} was generalized to
$U(N)$ in the unpublished work \cite{mmu} (see also \cite{dt}), and the matrix
integral representation was rederived recently in \cite{beasley,kallen} by a
direct localization of the path integral. We show that the spectral curve of
this matrix model agrees with our natural proposal for the B-model geometry. Since this curve is a 
symplectic transformation of the resolved conifold geometry, and since symplectic transformations do not 
change the $1/N$ expansion of the partition function \cite{eo,eotwo}, our result 
explains the empirical observation of \cite{dt,beasley} that the 
partition functions of the matrix models for different torus knots are all equal to the partition function of Chern--Simons theory on $\IS^3$ (up to an unimportant framing factor). 

This paper is organized as follows. In Section~2 we review the construction of
knot operators in Chern--Simons theory, following mainly \cite{llr}. In
section 3 we focus on the B-model point of view on knot invariants. We briefly
review the results of \cite{akv} on framed knots, and we show that a general
${\rm Sl}(2,\IZ)$ transformation of the spectral curve provides the needed framework to incorporate torus knots. This leads to a spectral curve for torus knots and links, which we analyze in detail. We compute some of the invariants with the topological recursion and we show that they agree with the knot theory counterpart. Finally, in Section~4 we study the matrix model representation of torus knots and we show that it leads to the spectral curve proposed in Section~3. We conclude in Section~5 with some implications of our work and prospects for future investigations. In the Appendix we derive the loop equations satisfies by the torus knot matrix model.

\sectiono{Torus knots in Chern--Simons theory}

First of all, let us fix some notations that will be used in the paper. We will denote by 
 \be
 \CU_\CK= {\rm P}\, \exp \oint_{\CK} A
 \ee
 the holonomy of the Chern--Simons connection $A$ around an oriented knot $\CK$, and by 
 \be
  \CW_R^{\cal K}=\tr_R\,  \CU_\CK
  \ee
  the corresponding Wilson loop operator in representation $R$. Its normalized vev will be denoted by 
 \be
 \label{wilsonvev}
 W_R(\CK)=\left \langle \tr_R \left( {\rm P}\, \exp \oint_{\CK} A\right)\right \rangle.
 \ee
 In the $U(N)$ Chern--Simons theory at level $k$, these vevs can be calculated in terms of the variables \cite{wittencs}
 \be
 \label{csvars}
 q=\exp\left( {2 \pi \ri \over k+N}\right), \qquad c=q^{N/2}.
 \ee
 When $R=\tableau{1}$ is the fundamental representation, (\ref{wilsonvev}) is related to the HOMFLY polynomial $\CH(\CK)$ of the knot $\CK$ as \cite{wittencs} 
 \be
 W_{\tableau{1}} (\CK)={c-c^{-1} \over q^{1/2} -q^{-1/2}} \CH(\CK). 
 \ee
Finally, we recall as well that the HOMFLY polynomial of a knot $\CH(\CK)$ has the following structure (see for example \cite{lickorish}), 
\be
\label{HOMFLYst}
\CH(\CK)=\sum_{i\ge 0} p_i (c^2) z^{2i}, \qquad z=q^{1/2}-q^{-1/2}. 
\ee

 Torus knots and links have a very explicit description \cite{llr} in the context of Chern--Simons gauge theory \cite{wittencs}. This description makes manifest the 
 natural ${\rm Sl}(2, \IZ)$ action on the space of torus knot operators, and it implements it in the quantum theory. It shows in particular that all torus knots can be obtained from the trivial knot or unknot by an ${\rm Sl}(2, \IZ)$ transformation. We now review the construction of torus knot operators in Chern--Simons theory, referring to \cite{llr} for more details. 
 
 Chern--Simons theory with level $k$ and gauge group $SU(N)$ can be canonically quantized on three-manifolds of 
 the form $\Sigma \times \IR$, where $\Sigma$ is a 
%orientable 
Riemann surface \cite{wittencs}. The resulting 
 Hilbert spaces can be identified with the space of conformal blocks of the $U(N)$ Wess--Zumino--Witten theory at level $k$ on $\Sigma$. When $\Sigma=\IT^2$ has genus one, 
 the corresponding wavefunctions can be explicitly constructed in terms  of theta functions on the torus \cite{bn,lr,emss,adpw}. The relevant theta functions are defined as 
\be
\Theta_{l,p}(\tau,a)=\sum_{\nu \in \Lambda_{\rm r}} \exp\left[ \ri \pi \tau l \left( \nu + {p \over l}\right)^2+ 2 \pi \ri l \left( \nu + {p \over l}\right) \cdot a\right], 
\ee
where $\tau$ is the modular parameter of the torus $\IT^2$, $\Lambda_{\rm r}$ is the root lattice of $SU(N)$, $a \in \Lambda_{\rm r} \otimes \IC$ and $p \in \Lambda_{\rm w}$, the weight lattice. Out of these theta functions we define the 
function
\be
\psi_{l,p}(\tau, a) =\exp \left( { \pi l \over 2 {\rm Im}\, \tau} a^2 \right) \Theta_{l, p}(\tau, a). 
\ee
Notice that, under a modular $S$-transformation, $a$ transforms as 
\be
a \rightarrow a/\tau. 
\ee
A basis for the Hilbert space of Chern--Simons theory on the torus is given by the Weyl antisymmetrization of these functions, 
\be
\label{wavef}
\lambda_{l, p}(\tau, a) =\sum_{w \in \CW} \epsilon(w) \psi_{l,w(p)}(\tau, a), 
\ee
where $\CW$ is the Weyl group of $SU(N)$, and 
\be
l=k + N.
\ee
The only independent wavefunctions obtained in this way are the ones where $p$ is in the fundamental chamber $\CF_l$, and they are in one-to-one correspondence 
with the integrable representations of the affine Lie algebra associated to $SU(N)$ with level $k$. We recall that the fundamental chamber
${\mathcal F}_l$ is given by $\Lambda_{\rm w}/l
\Lambda_{\rm r}$, modded out by the action of the Weyl group. For example,
in $SU(N)$ a weight $p=\sum_{i=1}^r p_i \lambda_i$ is in ${\mathcal F}_l$ if
\begin{equation}
\sum_{i=1}^r p_i < l,\,\,\,\,\,\, {\rm and} \,\,\ p_i >0, \, i=1, \cdots, r
\end{equation}
where $r=N-1$ is the rank of the gauge group. The wavefunctions (\ref{wavef}), where $p \in \CF_l$, span the Hilbert space $\CH_l(\IT^2)$ associated to Chern--Simons theory on $\IT^2$. 

%\FIGURE[ht]{
\begin{figure}[t]
\centering
\label{wilsonsolid}
\includegraphics[height=3cm]{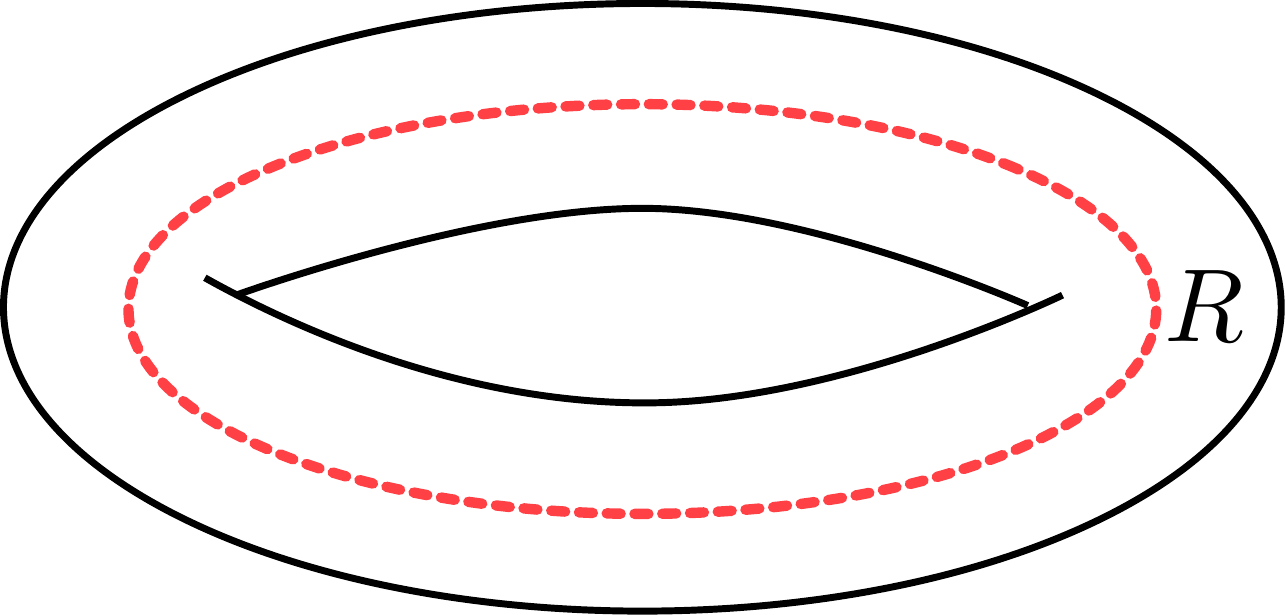} 
\caption{The path integral over a solid torus with the insertion of the Wilson line in representation $R$ gives the wavefunction defined in (\ref{wavef}).} 
%}
\end{figure}
The state described by the wavefunction $\lambda_{l,p}$ has a very simple representation in terms of path integrals in Chern--Simons gauge theory \cite{wittencs}. 
Let us write 
\be
p=\rho+\Lambda_R, 
\ee
where $\rho$ is the Weyl vector and $\Lambda_R$ is the highest weight associated to a representation $R$. Let us consider the path integral of Chern--Simons gauge theory on a solid torus $M_{\IT^2}$ with boundary $\partial M_{\IT^2} =\IT^2$, and let us insert a circular Wilson line 
\be
\CW_R^{(1,0)}=\tr_R \left( P\, \exp \oint_{\CK_{1,0}} A\right)
\ee
along the non-contractible cycle $\CK_{1,0}$ of the solid torus (see \figref{wilsonsolid}). This produces a wavefunction $\Psi(A)$, where $A$ is a gauge field on $\IT^2$. 
Let us now denote by $\omega(z)$ the normalized holomorphic Abelian differential on the torus, and let 
\be
H=\sum_{i=1}^R H_i \lambda_i 
\ee
where $H_i$, $\lambda_i$ are the Cartan matrices and fundamental weights of $SU(N)$, respectively. A gauge field on the torus can be parametrized as 
\be
A_z=\left( u_a \overline u \right)^{-1} \partial_z \left( u_a \overline u\right), \qquad A_{\bar z}=\left( u_a \overline u \right)^{-1} \partial_{\bar z} \left( u_a \overline u\right),
\ee
where 
\be
u:\IT^2 \rightarrow SU(N)^{\IC}
\ee
is a single-valued map taking values in the complexification of the gauge group, and 
\be
u_a =\exp\left( {\ri \pi \over {\rm Im}\, \tau} \int^{\bar z} {\overline {\omega(z')}} a\cdot H- {\ri \pi \over {\rm Im}\, \tau} \int^{z} \omega(z') \overline a\cdot H\right). 
\ee
In this way, the gauge field is written as a complexified gauge transformation of the 
complex constant connection 
\be
\sum_{i=1}^r a_i H_i.
\ee
After integrating out the non-zero modes of the gauge connection \cite{emss,llr}, one obtains an effective quantum mechanics problem where wavefunctions depend only on $a$, and they are given precisely by (\ref{wavef}). In particular, the empty solid torus corresponds to the trivial representation with $\Lambda_R=0$, 
and it is described by the ``vacuum" state
\be
\label{vacs}
\lambda_{l, \rho}. 
\ee
We will also represent the wavefunctions (\ref{wavef}) in ket notation, as $|R\rangle$, and the vacuum state (\ref{vacs}) will be denoted by $|0\rangle$. 

\begin{figure}
\begin{center}
\includegraphics[height=3.5cm]{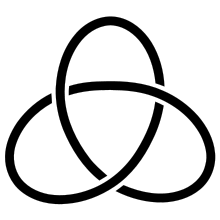} \qquad  \qquad  \qquad \includegraphics[height=3.5cm]{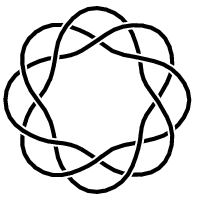}
\end{center}
\caption{The trefoil knot, shown on the left, is the $(2,3)$ torus knot. The knot shown on the right is the $(3,8)$ torus knot (these figures courtesy of {\it Wikipedia}).} 
\label{tknots}
\end{figure}

Torus knots can be defined as knots that can be drawn on the surface of a torus without self-intersections. They are labelled by two coprime integers $(Q,P)$, which 
represent the number of times the knot wraps around the two cycles of the torus, and we will denote them by $\CK_{Q,P}$. Our knots will be oriented, so the signs of $Q$, $P$ are relevant. We have a number of obvious topological equivalences, namely
\be
\label{sprops}
\CK_{Q,P} \simeq \CK_{P,Q}, \qquad \CK_{Q,P} \simeq \CK_{-Q, -P}. 
\ee
If we denote by $\CK^*$ the mirror image of a knot, we have the property 
\be
\label{tkmirror}
\CK^*_{Q,P}=\CK_{Q,-P}.
\ee
This means that, in computing knot invariants of torus knots, we can in principle restrict ourselves to knots with, say, $P>Q>0$. The invariants of the other torus knots can be 
computed by using the symmetry properties (\ref{sprops}) as well as the mirror property (\ref{tkmirror}), together with the transformation rule under mirror reflection
\be
\label{mirrorrule}
\left\langle \tr\,\CU_{\CK^*} \right \rangle(q,c)=\left\langle \tr\,\CU_{\CK} \right \rangle (q^{-1},c^{-1}).
\ee
All the knots $\CK_{1,f}$, with $f\in \IZ$, are isotopic to the trivial knot  or unknot. The simplest non-trivial knot, the trefoil knot, is the $( 2, 3)$ torus knot. It is depicted, together with the more complicated $(3,8)$ torus knot, in \figref{tknots}. 

Since torus knots can be put on $\IT^2$, a $(Q,P)$ torus knot in a representation $R$ should lead to a state in $\CH_l (\IT^2)$. As shown in \cite{llr}, these states 
can be obtained by acting with a {\it knot operator} 
\be
{\bf W}_R^{(Q,P)}:\CH_l(\IT^2) \rightarrow \CH_l(\IT^2),
\ee
on the vacuum state (\ref{vacs}). If we represent the states as wavefunctions of the form (\ref{wavef}), torus knot operators can be explicitly written as \cite{llr}
\be
\label{tkops}
{\bf W}_R^{(Q,P)}=\sum_{\mu \in M_R} \exp\left( -{\pi \over {\rm Im}\,\tau} \left( Q \overline \tau + P\right) a\cdot \mu + {Q \tau + P \over l }\mu\cdot {\partial \over \partial a} \right), 
\ee
where $M_R$ is the space of weights associated to the representation $R$. 
In the above description the integers $(Q,P)$ do not enter in a manifestly symmetric way, since 
$Q$ labels the number of times the knot wraps the non-contractible cycle of
the solid torus, and $P$ labels the number of times it wraps the contractible
cycle. However, knot invariants computed from this operator are symmetric in
$P$, $Q$; this is in fact a feature of many expressions for quantum invariants of torus knots, 
starting from Jones' computation of their HOMFLY polynomials in \cite{jones}. From (\ref{tkops}) one finds, 
\be
{\bf W}_R^{(Q,P)}\lambda_{l, p} =\sum_{\mu \in M_R} \exp\left[ \ri \pi \mu^2 {P Q \over l} + 2 \pi \ri  {P  \over l }p\cdot \mu \right] \lambda_{l, p+ Q\mu}. 
\ee

The torus knot operators (\ref{tkops}) have many interesting properties, described in detail in \cite{llr}. 
First of all, we have the property 
\be
{\bf W}_R^{(1,0)}\lambda_{l, \rho}=\lambda_{l, \rho+\Lambda_R},
\ee
which is an expected property since the knot $\CK_{1,0}$ leads to the Wilson line depicted in \figref{wilsonsolid}. 
Second, they transform among themselves under the action of the modular group of the torus ${\rm Sl}(2, \IZ)$. One finds \cite{llr}
\be
\label{modops}
M {\bf W}_R^{(Q,P)} M^{-1} ={\bf W}_R^{(Q,P)M}, \qquad M\in {\rm Sl}(2, \IZ),
\ee
where $(Q,P)M$ is the natural action by right multiplication. Since the torus knot $(Q,P)=(1,0)$ is the trivial knot or unknot, we conclude that a generic torus knot operator can be obtained by acting with an ${\rm Sl}(2,\IZ)$ transformation on the trivial knot operator. Indeed, 
\be
\label{modtknot}
M_{Q,P}{\bf W}_R^{(1,0)} M_{Q,P} ^{-1} = {\bf W}_R^{(Q,P)}, 
\ee
where $M_{Q,P}$ is the ${\rm Sl}(2, \IZ)$ transformation
\be
\label{QPmod}
M_{Q,P}=\begin{pmatrix} Q & P \\ \gamma & \delta \end{pmatrix}
\ee
and $\gamma$, $\delta$ are integers such that 
\be
\label{symcond}
Q\delta-P\gamma=1. 
\ee
Since $P, Q$ are coprime, this can be always achieved thanks to B\'ezout's lemma. 

\begin{figure}[t]
\centering
\label{torusframing}
\includegraphics[height=3cm]{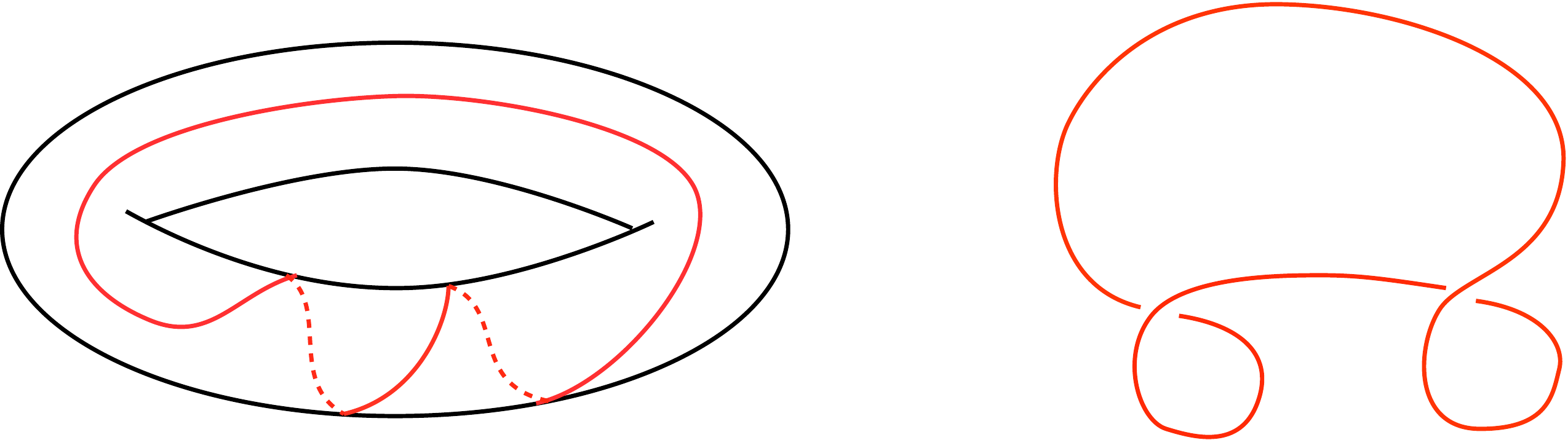} 
\caption{The knot operator with labels $(1,f)$ creates a Wilson line which winds once around the noncontractible cycle of the solid torus, and $f$ times around the 
contractible cycle. This corresponds to an unknot with $f$ ``ribbons," i.e. to an unknot with $f$ units of framing.} 
%}
\end{figure}

The final property we will need of the operators (\ref{tkops}) is that they make it possible to compute the vacuum expectation values of Wilson lines associated to torus knots in $\IS^3$. In fact, to construct a torus knot in $\IS^3$ we can start with an empty solid torus, act with the torus knot operator (\ref{tkops}) to create a torus knot on its 
surface, and then glue the resulting geometry to another empty solid torus after an $S$ transformation. We conclude that
\be
\label{knotinv}
W_R \left( \CK_{Q,P} \right)= { \langle 0 | S {\bf W}_R^{(Q,P)} |0\rangle \over  \langle 0 | S |0\rangle}, 
\ee
where we have normalized by the partition function of $\IS^3$. When performing this computation we have to remember that Chern--Simons theory produces invariants of {\it framed} knots \cite{wittencs}, and that a change of framing by $f$ units is implemented as
\be
W_R \left( \CK\right)  \rightarrow \re^{  2 \pi \ri f h_R}  W_R \left( \CK \right)
\ee
where
\be
h_R={\Lambda_R \cdot(\Lambda_R + 2 \rho) \over 2(k+N)}.
\ee
For knots in $\IS^3$ there is a standard framing, and as noticed already in \cite{llr}, torus knot operators naturally implement a framing of $QP$ units, as compared to the standard framing. For example, the knot operator 
\be
{\bf W}_R^{(1,f)}, \qquad f\in \IZ, 
\ee
creates a trivial knot but with $f$ units of framing \cite{llr,ilr}, see \figref{torusframing}. As we will see, the same natural framing $QP$ appears in the B-model for torus knots 
and in the matrix model representation obtained in \cite{lr,mmu,beasley}. 

The vev (\ref{knotinv}) can be computed in various ways, but the most efficient one was presented in \cite{stevan} and makes contact with the general formula for these invariants due to Rosso and Jones \cite{rj}. One first considers the knot operator 
\be
{\bf W}_R^{(Q,0)}
\ee
which can be regarded as the trace of the $Q$-th power of the holonomy around $\CK_{1,0}$. It should then involve the Adams operation on the representation ring, which 
expresses a character of the $Q$-th power of a holonomy in terms of other characters, 
\be
\label{adams}
{\rm ch}_R(U^Q)=\sum_V c^V_{R, Q} {\rm ch}_V(U). 
\ee
Indeed, one finds \cite{stevan}
\be
{\bf W}_R^{(Q,0)}=\sum_{V} c^{V}_{R, Q} {\bf W}_V^{(1,0)}.
\ee
If we introduce the diagonal operator \cite{stevan}
\be
{\bf D}_{P/Q} |R\rangle = \re^{2 \pi \ri {P/Q} h_R} |R\rangle
\ee
we can write an arbitrary torus knot operator as 
\be
\label{fracko}
{\bf W}_R^{(Q,P)}={\bf D}_{P/Q} {\bf W}_R^{(Q,0)}{\bf D}_{P/Q}^{-1}=T^{P/Q} {\bf W}_R^{(Q,0)} T^{-P/Q}, 
\ee
where 
\be
T^{P/Q} =\begin{pmatrix} 1 & P/Q \\ 0 & 1 \end{pmatrix}
\ee
is a ``fractional twist," in the terminology of \cite{mortonm}. The above identity can be interpreted by saying that the holonomy creating a $(Q,P)$ torus knot is equivalent to the $Q$-th power of the holonomy of a trivial knot, together with a {\it fractional} framing $P/Q$ (implemented by the operator ${\bf D}_{P/Q}$). As we will see, the same description arises in the B-model description of torus knots.  Since 
\be
{ \langle 0 | S |R\rangle \over  \langle 0 | S |0\rangle}={\rm dim}_q(R), 
\ee
the quantum dimension of the representation $R$, we find from (\ref{fracko}) that the vev (\ref{knotinv}) is given by
\be
\label{finaltorus}
W_R \left( \CK_{Q,P} \right) =\sum_V c^V_{R, Q} \re^{2 \pi \ri Q/P h_V} {\rm dim}_q(V).
\ee
This is precisely the formula obtained by Rosso and Jones in \cite{rj} (see also \cite{lz} for a more 
transparent phrasing). As pointed out above, in this formula the torus knot comes with a natural framing of $QP$ units. 

The formalism of torus knot operators can be also used to understand torus links. 
When $Q$ and $P$ are not coprime, we have instead a link $\CL_{Q,P}$ with $L={\rm gcd}(Q,P)$ 
components. From the point of view of the above formalism, the operator creating such a link can be obtained \cite{ilr,lm} 
by considering the product of $L$ torus knot operators with labels $(Q/L, P/L)$, i.e. 
\be
{\bf W}_{R_1, \cdots, R_L}^{(Q,P)} =\prod_{j=1}^L {\bf W}_{R_j}^{(Q/L,P/L)}. 
\ee
As explained in \cite{lmv}, this can be evaluated by using the fact that the torus knot operators provide a representation of the fusion rules of the affine Lie algebra \cite{llr}, therefore we 
can write
\be
{\bf W}_{R_1, \cdots, R_L}^{(Q,P)} =\sum_{R_s} N_{R_1, \cdots, R_L}^{R_s}  {\bf W}_{R_s}^{(Q/L,P/L)}, 
\ee
where the coefficients in this sum are defined by 
\be
R_1\otimes \cdots \otimes R_L =\sum_{R_s} N_{R_1, \cdots, R_L}^{R_s} R_s
\ee
and can be regarded as generalized Littlewood--Richardson coefficients. The problem of torus links reduces in this way to the problem of torus knots. Notice that in this formalism each component of the torus link has a natural framing $QP/L^2$. 

\sectiono{The B-model description of torus knots}

\subsection{Preliminaries}

Before discussing the B-model picture, we will recall the standard dictionary relating the correlators obtained in the knot theory side with the generating functions discussed in the B-model (see, for example, Appendix A in \cite{bkmp}). In the knot theory side we consider the generating function
\be
\label{totalopenf}
F(\CV)=\log \, Z(\CV), \qquad Z(\CV)=\sum_R W_R(\CK) \, \tr_R\, \CV
\ee
where $\CV$ is a $U(\infty)$ matrix, and we sum over all the irreducible representations $R$ (starting with the trivial one). It is often convenient to write the free energy $F(V)$ in terms of connected amplitudes in the basis 
labeled by vectors with nonnegative entries ${\bf k}=(k_1, k_2, \cdots)$. In this basis, 
\be
F(\CV)=\sum_{\bf k}{1 \over z_{\bf k}} W^{(c)}_{\bf k} \Upsilon_{\bf k}(\CV) 
\label{convev}
\end{equation}
where
\be
\Upsilon_{\bf k}(\CV) =\prod_{j=1}^{\infty} ({\rm Tr} \CV^j)^{k_j}, 
\qquad z_{\bf k}=\prod_j k_j! j^{k_j}.
\ee
The functional (\ref{totalopenf}) has a well-defined genus expansion, 
\be
F(\CV)=\sum_{g=0}^{\infty}\sum_{h=1}^{\infty} g_s^{2g-2+h} A_h^{(g)} (z_1, \cdots, z_h).
\ee
In this equation, $g_s$ is the string coupling constant (\ref{gscoupling}), and we have written
\be
\label{idenwords}
 {\rm Tr}\, \CV^{w_1} \cdots \tr \, \CV^{w_h} \leftrightarrow m_w (z)=\sum_{\sigma \in S_h} \prod_{i=1}^h z_{\sigma(i)}^{w_i}
 \ee
where $m_w(z)$ is the monomial symmetric polynomial in the $z_i$ and $S_h$ is the symmetric group of $h$ elements. After setting $z_i=p_i^{-1}$, the functionals 
$A_h^{(g)} (z_1, \cdots, z_h)$ are given by 
\be
A_h^{(g)} (p_1, \cdots, p_h)=\int \rd p_1 \cdots \rd p_h W_{g,h}(p_1, \cdots, p_h),
\label{eq:wgh}
\ee
where the functionals $W_{g,h}$ are the ones appearing naturally in the B-model through the topological recursion. 

\subsection{Symplectic transformations in the resolved conifold}

We now briefly review the B-model description of the framed unknot proposed in \cite{akv}. 

According to the Gopakumar--Vafa large $N$ duality and its extension to Wilson loops in \cite{ov}, knot and link invariants are dual to open topological 
string amplitudes in the resolved conifold 
\be
\CO(-1) \oplus \CO(-1) \rightarrow \IP^1 
\ee
with boundary conditions set by Lagrangian A-branes. We recall the basic dictionary of \cite{gv}: the string coupling constant $g_s$ is identified 
with the renormalized Chern--Simons coupling constant, 
\be
\label{gscoupling}
g_s ={2\pi \ri \over k+N}, 
\ee
while the K\"ahler parameter of the resolved conifold is identified with the 't Hooft parameter of $U(N)$ Chern--Simons theoy, 
\be
\label{thooft}
t={2 \pi \ri N \over k+N} =g_s N. 
\ee
The unknot and the Hopf link correspond to toric A-branes of the type
introduced in \cite{av, ov} and their Chern--Simons invariants can be computed in the dual A-model 
picture by using localization \cite{kl} or the topological vertex \cite{akmv,ikp,zhou}.

By mirror symmetry, there should be a B-model 
version of the Gopakumar--Vafa large $N$ duality. We recall (see for example \cite{akv} and references therein) 
that the mirror of a toric Calabi--Yau manifolds is described by an algebraic curve in $\IC^* \times \IC^*$ (also 
called spectral curve) of the form 
\be
\label{hcurve}
H\left(\re^u, \re^v\right)=0. 
\ee
We will denote
\be
U=\re^u, \quad V=\re^v
\ee
The mirrors to the toric branes considered in \cite{av} boil down to points in this curve, and the disk amplitude for topological strings 
is obtained from the function $v(u)$ that solves the equation (\ref{hcurve}). Different choices of parametrization of this point lead to different 
types of D-branes, as we will discuss in more detail. According to the conjecture of \cite{mm,bkmp}, higher open string 
amplitudes for toric branes can be obtained by applying the topological recursion of \cite{eo} to the spectral curve (\ref{hcurve}). 

The mirror of the resolved conifold can be described by the spectral curve (see \cite{akv,bkmp}) 
\be
\label{rescurve}
H(U,V)=V-c^{-1} UV +c U-1=0, 
\ee
where
\be
\label{cdef}
c=\re^{t/2}.
\ee
By mirror symmetry, $t$ corresponds to the K\"ahler parameter of the resolved conifold. Due to the identification in (\ref{thooft}), the variable $c$ appearing in the spectral curve is identified with the Chern--Simons variable introduced in (\ref{csvars}). The mirror brane to the unknot with zero framing, $\CK_{1,0}$, is described by a point in this curve, parametrized by $U$, and the generating function of disk amplitudes 
\be
-\log V(U)= -\log \left( {1-c U \over 1-c^{-1} U} \right)=\sum_{n\ge 0} \langle \tr \, \CU_{\CK_{1,0}}^n\rangle_{g=0} U^n, 
\ee
can be interpreted as the generating function of planar one-point correlators for the unknot. 

As pointed out in \cite{akv}, in writing the mirror curve (\ref{hcurve}) there is an ambiguity in the choice of variables given by an ${\rm Sl}(2, \IZ)$ transformation, 
\be
\label{gensym}
\ba
X &=U^Q V^P,\\
Y &= U^\gamma V^\delta, 
\ea
\ee
where $Q,P, \gamma, \delta$ are the entries of the ${\rm Sl}(2, \IZ)$ matrix (\ref{QPmod}). However, only modular transformations of the form 
\be
M_{1,f}=\begin{pmatrix} 1 & f \\ 0 &1 \end{pmatrix}, \qquad f \in \IZ,
\ee
were considered in \cite{akv}. In the case of the mirror of the resolved conifold they were interpreted as adding $f$ units of framing to the unknot. It was 
argued in \cite{akv} that only these transformations preserve the geometry of the brane at infinity. The resulting curve can be described as follows. We first rescale the variables as
\be
\label{rescaling}
U, \, X \rightarrow c^f U, \, c^f X.  
\ee
The new curve is defined by, 
\be
\label{framedc}
\ba
X&=U \left( {1-c^{f +1} U \over 1-c^{f-1} U} \right)^f, \\
V&={1-c^{f +1} U \over 1-c^{f-1} U}, 
\ea
\ee
and as proposed in \cite{mmopen,bkmp}, the topological recursion of \cite{eo} applied to this curve gives all the Chern--Simons invariants of the framed unknot. 

The general symplectic transformation (\ref{gensym}) plays a crucial r\^ole in the formalism of \cite{eo,bkmp}, where it describes the group of symmetries associated to the 
closed string amplitudes derived from the curve (\ref{hcurve}). It is natural to ask what is the meaning of these, more general transformation. In the case of the resolved conifold, and in view of the modular action (\ref{modtknot}) on torus knot operators, it is natural to conclude that the transformation associated to the matrix $M_{Q,P}$ leads to the mirror brane to a torus knot. We will now give some evidence that this is the case. In the next section we will derive this statement from the matrix model 
representation of torus knot invariants. 

\subsection{The spectral curve for torus knots}

Let us look in some more detail to the general modular transformation (\ref{gensym}). We first redefine the $X, U$ variables as
\be
U\rightarrow c^{P/Q} U, \qquad X \rightarrow c^P X.
\ee
This generalizes (\ref{rescaling}) and it will be convenient in order to match the knot theory conventions. The first equation in (\ref{gensym}) reads now
\be
X = U^Q \left( {1-c^{P/Q +1} U \over 1-c^{P/Q-1} U} \right)^P
\ee
and it defines a multivalued function 
\be
\label{ux}
U=U(X)=X^{1/Q}+\cdots
\ee
Equivalently, we can define a local coordinate $\zeta$ in the resulting curve as 
\be
\label{zetaU}
\zeta\equiv X^{1/Q}=U \left( {1-c^{P/Q +1} U \over 1-c^{P/Q-1} U} \right)^{P/Q}. 
\ee
Combining (\ref{ux}) with the equation for the resolved conifold (\ref{rescurve}) we obtain a function $V=V(X)$. After re-expressing $U$ in terms of $X$ in the second equation of (\ref{gensym}), and using (\ref{symcond}), we find that the dependence of $Y$ on the new coordinate $X$ is of the form
\be
\label{logy}
\log Y ={\gamma \over Q} \log X +{1\over Q} \log \, V(X). 
\ee
The term $\log V(X)$ in this equation has an expansion in fractional powers of
$X$ of the form $n/Q$, where $n\in\IZ$.
%, and the power of $X$ is an integer 
%whenever $n$ is a multiple of $Q$. 
By comparing (\ref{zetaU}) to (\ref{framedc}), we conclude that 
the integer powers of $X$ appearing in the expansion of $\log V(X)$ are the integer powers of $\zeta^Q$ in the curve (\ref{framedc}), but with fractional framing 
\be
f=P/Q.
\ee
This is precisely the description of $(Q,P)$ torus knots appearing in (\ref{fracko})! It suggests that the integer powers of $X$ 
in the expansion of $\log V(X)$ encode vevs of torus knot operators. Since the first term in (\ref{logy}) is not analytic at $\zeta=0$, we can regard $\log Y$ (up to a factor 
of $Q$) as the spectral curve describing torus knots in the
B-model. Equivalently, if we want a manifestly analytic function of $\zeta$ at
the origin, as is the case in the context of the matrix model describing torus knots, we can consider the spectral curve in the $(X, V)$ variables defined by
\be
\label{tksc}
\ba
X&=U^Q \left( {1-c^{P/Q +1} U \over 1-c^{P/Q-1} U} \right)^P, \\
V&=  {1-c^{P/Q +1} U \over 1-c^{P/Q-1} U}. 
\ea
\ee
This curve can be also written as 
\be
\label{polcurve}
H_{Q,P}(X, V)=V^P (V-1)^Q -c^{P-Q} X (V-c^2)^Q=0.
\ee
Notice that, when $Q=1$, $P=0$ (i.e. for the unknot with zero framing) we recover the standard equation (\ref{rescurve}) for the resolved conifold, and for $Q=1$, $P=f$ 
we recover the curve of the framed unknot (\ref{framedc}). In the curve (\ref{tksc}), $X$ is the right local variable to expand in order to obtain the invariants. The topological 
recursion of \cite{eo}, applied to the above curve, leads to generating functionals which can be expanded in powers of $X^{1/Q}$ around $X=0$. The coefficients 
of the integer powers 
of $X$ in these expansions give the quantum invariants of the $(Q,P)$ torus knot, in the $QP$ framing.

When $Q$ and $P$ are not coprime, the above curve describes a torus link with $L={\rm gcd}(Q,P)$ components. Up to a redefinition of the local variable of the curve, 
the disk invariants have the same information of the disk invariants of the $(Q/L, P/L)$ torus knot. However, as we will see in a moment, the $L$-point functions 
obtained from the topological recursion compute invariants of the torus link.

\subsection{One-holed invariants}
\subsubsection{Disk invariants}
The simplest consequence of the above proposal is that the integer powers of $X$ in the expansion of $-\log V(X)$ give the invariants 
\be
\left\langle \tr\,\CU^m_{\CK_{(Q,P)}} \right \rangle_{g=0}.
\label{eq:trun}
\ee
 We will now compute in closed form the generating function $-\log V(X)$. The equation (\ref{zetaU}) defines the local coordinate $\zeta$ as a function of $U$, and it can be easily inverted (by using for example Lagrange inversion) to give, 
\be
U=\sum_{n=1}^{\infty}a_n \zeta^n, 
\ee
where $a_1=1$ and 
\be
a_n={P\over Q} {c^{ (n-1) P/Q} \over (n-1)!} \sum_{k=0}^{n-1} (-1)^k {n-1 \choose k} \prod_{j=-n+k+2}^{k-1} \left( {P n \over Q}-j\right) c^{n-1-2k}, \qquad n\ge 2.
\ee
This is essentially the result obtained in \cite{mv}, eq. (6.6), in the context of framed knots, but with a fractional framing $P/Q$. From this expansion it is easy to obtain
\be\label{eq:defWnc}
-\log V(X)=\sum_{n\ge 1} W_n(c) X^{n/Q}, 
\ee
where
\be
\label{wfund}
W_n(c)={1\over n!} \sum_{\ell=0}^n (-1)^{n+\ell} {n\choose \ell} c^{2\ell + n (P/Q-1)} \prod_{j=-\ell+1}^{n-1-\ell} \left( {n P \over Q}-j\right), 
\ee
which is again essentially the result obtained in eq. (6.7) of \cite{mv}. Integer powers of $X$ corresponds to $n=Q m$, $m \in \mathbb{N}$, and we
conclude that the the planar limit of \eqref{eq:trun} for the $(Q,P)$ torus knot with framing $QP$ should be given by
\ben
\left\langle \tr\,\CU^m_{\CK_{Q,P}} \right \rangle_{g=0} &=& {1\over m Q!}
\sum_{\ell=0}^{mQ} (-1)^{mQ+\ell} {mQ\choose \ell} c^{2\ell +
  m(P-Q)}\prod_{j=-\ell+1}^{mQ-1-\ell} \left(mP-j\right) \nn \\
&=&  \frac{(-1)^{mQ} c^{m (P-Q)} (m P-1)!}{(m P - m Q )! (m Q)!} \, _2F_1\left(m P,-m Q;m P-m
   Q+1;c^2\right).
\label{leading}
\een
This can be verified for the very first values of $Q,P$. For example, we obtain:
\be
\ba
\left\langle \tr\,\CU_{\CK_{2,3}} \right \rangle_{g=0}&=c-3c^3+ 2 c^5, \\
\left\langle \tr\,\CU_{\CK_{2,5}} \right \rangle_{g=0}&=c^3 \left(3 c^4-5 c^2+2\right),\\
\left\langle \tr\,\CU_{\CK_{3,5}} \right \rangle_{g=0}&=c^2 \left(7 c^6-15 c^4+10 c^2-2\right),
\ea
\ee
which give the correct result for the genus zero knot invariants. In particular, the above expression turns out to be symmetric under the exchange 
of $Q$ and $P$, although this is not manifestly so. 

The expression (\ref{leading}) can be written in various equivalent ways, and it is closely related to a useful knot invariant. Indeed, the vev 
\be
\left\langle \tr\,\CU_{\CK} \right \rangle_{g=0}(c)
\ee
is, up to an overall factor of $c-c^{-1}$, the polynomial $p_0(c^2)$ appearing in the expansion (\ref{HOMFLYst}). This polynomial 
plays a distinguished r\^ole in knot theory, and this seems to be closely related to the fact that it is the leading term in the large $N$ expansion (this was first pointed out in 
\cite{guada}). The polynomial $p_0(c^2)$ of torus knots appears in the work of Traczyk \cite{traczyk} on periodicity properties of knots, but a closed expression as 
a function of $Q,P$ does not seem 
to be available in the literature. Using the above results, and performing various simple manipulations, we find the following expression, valid for $Q,P>0$:
\be
\label{leadingkc}
p_0^{\CK_{(Q,P)}}(c^2)=c^{(P-1)(Q-1)}  {(P+Q-1)! \over P! Q!} {~}_2F_1\left( 1-P, 1-Q, 1-P-Q;c^2\right).
\ee
Here ${~}_2F_1\left( a, b, c;x\right)$ is the standard Gauss' hypergeometric function. 
Of course, since the indices are negative, the r.h.s. is a polynomial in $c^2$.  
 In writing (\ref{leadingkc}), which is manifestly symmetric under the exchange of 
$P$ and $Q$, we have implemented two small changes w.r.t. (\ref{leading}). 
First of all, invariants of knots in $\IS^3$ are usually presented in the standard framing, 
while the results obtained for the spectral curve correspond to a torus knot with framing $QP$. In order to 
restore the standard framing we have to multiply the expression (\ref{leading}) by $c^{-PQ}$. Second, 
our labeling of the torus knot does not agree with the standard conventions in the 
literature: what we call the $(Q,P)$ torus knot is usually regarded as a $(Q, -P)$ torus knot. 
This means that we have to apply the mirror transformation (\ref{mirrorrule}) to our invariant, which 
implies in particular that
\be
\label{mirrorlead}
p_0^{\CK^*}(c^2)=  p_0^{\CK}(c^{-2}). 
\ee
After implementing these changes, one obtains (\ref{leadingkc}) from (\ref{leading}). Of course, if $(Q,P)$ are not both positive or both negative, we can use (\ref{mirrorlead}) to compute the invariant. 

The spectral curve (\ref{tksc}) gives, on top of the invariants of torus knots, information about other invariants associated to the torus knot, encoded in the 
coefficients of the fractional powers of $X$. They correspond to fractional
powers of the holonomy around the knot. As we will see in Section \ref{sec:mm} 
these invariants have a natural interpretation in the matrix model for torus knots. 

\subsubsection{All-genus invariants}
Even more remarkably, the close relation of the invariants of the ($P$, $Q$)
torus knots to the ones of the unknot at fractional framing can be further
pushed to derive an {\it all-genus} completion of \eqref{leadingkc} in terms
of $q$-hypergeometric polynomials. To see this, notice that one-holed
invariants at winding number $m$ receive contributions from vevs in hook
representations $R_{r,s}$ 
\ben
\bra \tr\,\CU^m_{\CK} \ket &=& \sum_R \chi_R(k_m) 
%q^{\kappa_R  p/2}c^{\ell(R)}
W_R(\CK) = \sum_{R_{m,s}} (-1)^s 
%q^{m(m-1)-2 m s}c^{pm} 
W_{R_{m,s}}(\CK)
\een
where $k_m$ is the conjugacy class of a length $m$ cycle in $S_m$, and
$R_{r,s}$ denotes a hook representation with $s+1$ rows. For the framed unknot, we have that 
\be
\bra \tr\,\CU^m_{\CK_{1,f}} \ket =\sum_{R_{m,s}} (-1)^s q^{2 \pi \ri f h_{R_{m,s}}} {\rm dim}_q(R_{m,s}). 
\ee
The quantum dimension of the representation $R_{m.s}$ can be written as 
\be
{\rm dim}_q(R_{m,s})= \frac{q^{m(m-1)/4-s m /2}}{[m][m-s-1]![s]!} c^{2m}
\prod_{i=1}^{m-s}\left(1-\frac{1}{q^{i-1} c^2}\right)   \prod_{i=1}^s  \left(1-\frac{q^i}{c^2}\right)
\ee
where for $n \in \mathbb{N}$ the $q$-number $[n]$ and the $q$-factorial $[n]!$ are defined as
\ben
[n]=q^{n/2}-q^{-n/2}, \qquad [n]!=[n][n-1]\dots[1].
\een
Upon applying the Cauchy binomial formula
\be
\sum_{s=0}^m t^s q^{s(m+1)/2} {m \brack s}=\prod_{j=1}^m(1+t q^j)
\ee 
we obtain the finite sum 
\be
\bra \tr\,\CU^m_{\CK_{1,f}} \ket =  \sum_{\ell=0}^m c^{2\ell +m f -m} (-1)^{m+\ell}  {1\over  [m-\ell]! [\ell]!} {[m f+\ell-1]! \over [mf-m+\ell]!},
\label{eq:qfunk}
 \ee
for the framed unknot at winding number $m$ with $f$ units of framing, which can be regarded as a $q$-deformed version of the formulae of \cite{mv} for the framed disc.
Following exactly the same line of reasoning as we did for the planar case, the full un-normalized HOMFLY polynomial for ($Q$, $P$) torus knots is obtained from \eqref{eq:qfunk} upon sending $f\to P/Q$, $m\to Q$:
\ben
\bra \tr\,\CU_{\CK_{Q,P}} \ket &=& \sum_{\ell=0}^Q (-1)^{Q+\ell}c^{2\ell +P
  -Q} {1\over  [Q-\ell]! [\ell]!}  {[P+\ell-1]! \over [P-Q+\ell]!} \nn \\
&=&  \frac{(-1)^{Q} c^{P-Q} [P-1]!}{[P -  Q ]! [Q]!} \, _2\phi_1\left(P,- Q; P- Q+1;c^2\right),
\label{qleading}
\een
where the $q$-analogue of Gauss' hypergeometric function is defined by
\be
_2\phi_1\left(a, b, c; q, z\right)=\sum_{n=0}^{\infty} {(a;q)_n (b:q)_n \over (q;q)_n (c;q)_n} z^n,
\ee
and the $q$-Pochhammer symbol is given as $(a,q)_n = [a+n-1]!/[a-1]!$ . Upon taking the $q\to 1$ limit, we recover \eqref{leading}. The natural
$q$-extension of \eqref{leadingkc} leads to the following expression for the HOMFLY polynomial of a torus knot,
\be
{1\over q^{1/2} -q^{-1/2}} \CH(\CK_{Q,P})=c^{(P-1)(Q-1)} {[P+Q-1]! \over [P]! [Q]!}  \, {~}_2 \phi_1\left(1-P, 1-Q, 1-P-Q;q, c^2\right).
\label{qleadingkc}
\ee
Again, since $P>0$ and $Q>0$, the series truncates to a degree $d=\mathrm{min}(P-1,Q-1)$ polynomial in $c^2$. It can be also written as 
\be
{1\over q^{1/2} -q^{-1/2}} \CH(\CK_{Q,P})=c^{(P-1)(Q-1)} {1\over [P] [Q]} \sum_{k=0}^{d} {[P+Q-k-1]! \over [P-k-1]! [Q-k-1]! [k]!} (-1)^k c^{2k}.
\ee
which is the result obtained in \cite{gorsky} for the HOMFLY polynomial of a torus knot. 

With \eqref{qleadingkc} at hand we can straightforwardly extract the higher genus corrections to \eqref{leadingkc}. Expanding the $q$-factorials around $q=1$
\ben
[n]! &=& (-1)^n n!  (1-q)^n q^{-\frac{1}{4} n (n+1)} \Bigg(1+\frac{1}{4} \left(n^2-n\right) (q-1) + \bigg(\frac{1}{18} (n-2) (n-1) n+  \nn \\ &+& \frac{1}{96} (n-2) (n-1) (3 n-1) n\bigg) (q-1)^2+\mathcal{O}(q-1)^3\Bigg),
\een
we obtain for example the closed expression
\ben
p_1^{\mathcal{K}_{Q,P}}(c) &=& -\frac{c^2}{48}\frac{\rd^2 p_0^{\mathcal{K}_{Q,P}}(c)}{\rd c^2}
+\frac{c+3c^3}{48(1-c^2)} \frac{\rd p_0^{\mathcal{K}_{Q,P}}(c)}{\rd c}
+\frac{1}{48} \left(P^2 \left(Q^2-1\right)-Q^2-3\right)p_0^{\mathcal{K}_{Q,P}}(c)
 \nn \\ &=& \frac{c^{(P-1) (Q-1)}(P+Q-1)!}{24 P! Q!} \Bigg(\frac{2 c^2 (P-1) P (Q-1) Q}{P+Q-1} \,   _2F_1\left(2-P,2-Q;-P-Q+2;c^2\right)  \nn \\ 
&+& (P^2 Q-P^2+P Q^2-P Q-Q^2-1) \, _2F_1\left(1-P,1-Q;-P-Q+1;c^2\right)\Bigg). \nn \\
\label{eq:p1}
\een
We get for instance
\ben
p_1^{\mathcal{K}_{2,3}}(c) &=& \frac{1}{12} c^2 \left(c^2+10\right), \\
p_1^{\mathcal{K}_{2,5}}(c) &=& -\frac{5}{12} c^4 \left(2 c^2-9\right), \\
p_1^{\mathcal{K}_{3,5}}(c) &=& \frac{5}{12} c^8 \left(2 c^4-32 c^2+49\right).
\een
in complete agreement with explicit computations using the Rosso--Jones formula \eqref{finaltorus}.
%which is the correct result for the knot invariants.

\subsection{Higher invariants from the spectral curve}

Let us now move to the case of higher invariants  by applying the
Eynard--Orantin recursion \cite{eo}  to the spectral data \eqref{tksc} or (\ref{gensym}). Let
$\Gamma_{Q,P}\simeq \mathbb{CP}^1$ be the projectivization of the affine curve
\eqref{polcurve}. We will take $U$ as an affine co-ordinate on $\mathbb{CP}^1$
and we will keep using $X, Y$ for the meromorphic extensions $X, Y: \Gamma_{Q,P}
\to \mathbb{CP}^1$ of \eqref{tksc}; we will finally call $\{q_i\}$ the quadratic
ramification points  of the $X\to\mathbb{CP}^1$ covering map.  Following
\cite{eo}, we recursively define a doubly infinite sequence of meromorphic
differentials $\omega_{g,h}(U_1, \cdots, U_h)  \rd X(U_1)\dots \rd X(U_h) \in
\mathcal{M}^h(\mathrm{Sym}^h(\Gamma_{Q,P}))$, $g\geq 0$, $h\geq 1$ on the $h^{\rm th}$ symmetric product of $\Gamma_{Q,P}$ as
\ben
%\omega_{0,1}(U) &=& \frac{\log{Y(U)}}{X(U)}, \\
\label{eq:02berg}
\omega_{0,2}(U_1, U_2) &=& B(U_1,U_2), \\
\omega_{g,h+1}(U_0, U_1 \ldots, U_h) &=& \sum_{q_i}  \underset{Z=q_i}{\rm Res~} K(U_0,U) \Big ( W_{g-1,h+2} (U, \overline{U}, U_1, \ldots, U_{h} )\nn \\
&+& {\sum_{l=0}^g}  {\sum'_{J\subset H}} W^{(g-l)}_{|J|+1}(U, U_J) W^{(l)}_{|H|-|J| +1} (\overline{U}, U_{H\backslash J}) \Big).
\label{eq:wghrec}
\een
In \eqref{eq:02berg}, $B(z, w) \rd z \rd w$ is the Bergmann kernel of $\Gamma_{Q,P}$, namely, the unique double differential with a double pole at $z=w$ and holomorphic elsewhere. Since $\Gamma_{P,Q}\simeq \mathbb{CP}^1$, it reads simply
\be
B(U_1, U_2) = \frac{1}{(U_1-U_2)^2}.
\label{eq:berg}
\ee
On the r.h.s. of \eqref{eq:wghrec}, $\overline{U}$ is the conjugate point to $U$ near $U=q_i$ under the $X$ projection (i.e. $X(U)=X(\overline{U})$, $U\neq \overline{U}$), the recursion kernel $K(U_1,U_2)$ is defined as
\be
K(U_1,U_2) = -\frac{X(U_2)}{2 X'(U_2)} \frac{\int_{\overline{U}_2}^{U_2} B(U_1,U') \rd U'}{\log{Y(U_2)}-\log{Y(\overline{U}_2)}},
\ee
with $I \cup J=\{U_1,\dots, U_h\}$, $I \cap J=\emptyset$, and $\sum'$ denotes omission of the terms $(h,I)=(0,\emptyset)$ and $(g,J)$. \\

The identification of \eqref{tksc} as the spectral curve associated to ($P$, $Q$) torus knots in $\mathbb{S}^3$ entails the identification of the differentials $\omega_{g,h}(U(X_1), \cdots, U(X_h)) \rd X_1\dots \rd X_h$ with the connected generating functions $W_{g,h}(U(X_1), \cdots, U(X_h)) \rd X_1\dots \rd X_h$ of \eqref{eq:wgh} for all $(g,h)\neq (0,2)$; in the exceptional case $(g,h)=(0,2)$, the annulus function is obtained from the Bergmann kernel upon subtraction of the double pole in the $X$ co-ordinate
\be
W_{0,2}(X_1, X_2) = B(X_1,X_2)-\frac{1}{(X_1-X_2)^2}.
\label{eq:w02}
\ee

With \eqref{tksc} and \eqref{eq:02berg}-\eqref{eq:wghrec} at hand, it is
straightforward to apply the topological recursion to compute higher invariants for torus knots. For the annulus function we obtain from \eqref{eq:02berg} and \eqref{eq:w02} that
\be
W_{0,2}(U_1, U_2) = \frac{1}{(U_1-U_2)^2}-\frac{X'(U_1) X'(U_2)}{(X(U_1)-X(U_2))^2}.
\label{eq:w02ex}
\ee
The planar part of connected knot invariants \eqref{convev} in the conjugacy
class basis $W^{(c)}_{\mathbf{k}}$, where  $\sum_i
k_i=2$ for the annulus function, can then be straightforwardly computed as
\be
W^{(c)}_{\mathbf{k}}\Big|_{g=0}=\mathrm{Res}_{U_1=\infty}\mathrm{Res}_{U_2=\infty} X(U_1)^n X(U_2)^m  W_{0,2}(U_1,U_2) \rd U_1 \rd U_2,
\ee
with $k_i = \delta_{in}+\delta_{im}$. We find explicitly for $Q=2$
\ben
W^{(c)}_{(2,0,0,\dots)}\Big|_{g=0} &=& \frac{1}{4} (c^2-1) P c^{2 P-4}
\left(c^2 (P+1)-P+1\right) \nn \\ & \times &\left(c^4 (P+1) (P+2)-2 c^2 P^2+P^2-3 P+2\right) \nn \\ \\
W^{(c)}_{(1,1,0,0,\dots)}\Big|_{g=0} &=& \frac{1}{9} P c^{3 P-6} \Bigg[6 \left(c^{12}-1\right)+4 \left(c^2-1\right)^6 P^5+ 24 \left(c^2+1\right) \left(c^2-1\right)^5 P^4  \nn \\ &+& \left(55 c^4+82 c^2+55\right) \left(c^2-1\right)^4 P^3+12 \left(5
   c^6+8 c^4+8 c^2+5\right) \left(c^2-1\right)^3 P^2 \nn \\ &+&\left(31 c^8+44 c^6+48 c^4+44 c^2+31\right) \left(c^2-1\right)^2 P\Bigg]
\een
and for $Q=3$
\ben
W^{(c)}_{(2,0,0,\dots)}\Big|_{g=0} &=& \frac{1}{24} P c^{2 P-6} \Bigg(c^{12}
(P+1)^2 (P+2)^2 (P+3)-6 c^{10} P (P+1)^2 (P+2)^2  \nn \\ &+& 3 c^8 P (P+1)^2
(5 P (P+1)-4) 4 c^6 P \left(5 P^4-7 P^2+2\right) + 3 c^4 (P-1)^2 P  \nn
\\ & \times & (5 (P-1) P-4)-6 c^2 P
   \left(P^2-3 P+2\right)^2+(P-3) \left(P^2-3 P+2\right)^2\Bigg), \nn \\
\een
in agreement with the corresponding knot invariants; notice that the case of
torus links is also encompassed as soon as $\mathrm{gcd}(P,Q)>1$, with the
Hopf link invariants appearing as the $(P,Q)=(2,2)$ case. \\

To compute higher order generating functions we resort to \eqref{eq:wghrec}. The regular branch points are
\be
q_\pm = \frac{c^{\frac{P}{Q}-1} \left(\pm\sqrt{c^2-1} \sqrt{\left(c^2-1\right) P^2+2
   \left(c^2+1\right) P Q+\left(c^2-1\right) Q^2}+\left(c^2-1\right) P+c^2
   Q+Q\right)}{2 Q}
\ee
and as will see they are precisely the ramification points that lie on the physical sheet of the spectral curve. 
For the case $g=1$, $h=1$ we obtain
\ben
\omega_{1,1}(U) &=&
\left(c^2-1\right) \left(-c^{\frac{P}{3}+1}\right) \left(c^{2 P/3}-U^2\right) \Bigg(81 c^{\frac{4 P}{3}+2}+81 c^2 U^4-6 U^3 \Big(c^2 (P+3) (P (2 P+3) \nn \\ &+& 9) - (P-3) (P (2 P-3)+ 9)\Big)   c^{\frac{P}{3}+1}-6 U \Big(c^2 (P+3) (P (2 P+3)+9)-(P-3)\nn \\ & &  (P (2 P-3)+9)\Big) c^{P+1}+U^2 \left(c^4 (P+3)^4-2 c^2 \left(P^4-54 P^2-162\right)+(P-3)^4\right) \nn \\ & & c^{2 P/3}\Bigg)\Bigg/\Bigg(12 \left(3 c^{\frac{2 P}{3}+1}-U \left(c^2 (P+3)-P+3\right) c^{P/3}+3 c U^2\right)^4\Bigg) \nn \\
\een
and it is immediate to extract genus one, 1-holed knot invariants as
\be
W_{\mathbf{k}}\Big|_{g=1}=\mathrm{Res}_{U=\infty} X(U)^n \omega_{1,1}(U) \rd U,
\ee
where in this case $k_i=\delta_{in}$.
For example
\be
\begin{array}{rclr}
W_{(1,0,0,\dots)}\Big|_{g=1} &=& -\frac{1}{24} \left(c^2-1\right) c^{P-1} & \quad Q=1\\
W_{(1,0,0,\dots)}\Big|_{g=1} &=&\frac{1}{48} \left(c^2-1\right) c^{P-2} \left(c^2 (P+1)
(P (P+2)-4)+(1-P) ((P-2) P-4)\right) & \quad Q=2 \\
W_{(1,0,0,\dots)}\Big|_{g=1} &= & \frac{1}{144} c^{P-3} \Bigg(c^6 (P+1) (P+2) (2 P
(P+3)-9)-3 c^4 P (P+1) (2 P (P+1)   & \quad Q=3
\\ &-& 5) 3 c^2 (P-1) P (2 (P-1) P-5)+(2-P)
(P-1) (2 (P-3) P-9)\Bigg) 
\end{array}
\ee
which reproduce \eqref{eq:p1} at fixed $Q$. Similarly, higher winding
invariants can be found to reproduce the correct knot invariants.

\sectiono{The matrix model for torus knots}
\label{sec:mm}
In this section we study the matrix model representation for quantum, colored invariants of torus knots. We first give a derivation of the matrix model 
which emphasizes the connection to the Rosso--Jones formula (\ref{finaltorus}), and then we use standard techniques in matrix models to derive the 
spectral curve describing the planar limit of the invariants. 

\subsection{A simple derivation of the matrix model}

The colored quantum invariants of torus knots admit a representation in terms of an integral over the Cartan 
algebra of the corresponding gauge group. Such a representation was first proposed for $SU(2)$ in \cite{lr}, and then 
extended to simply-laced groups in \cite{mmu} (see also \cite{dt}). More recently, the matrix integral for torus knots was 
derived by localization of the Chern--Simons path integral \cite{beasley} (another localization procedure which leads to the same 
result has been recently proposed in \cite{kallen}). 

The result obtained in these papers reads, for any simply-laced group $G$,
\be
\label{tkmi}
W_R(\CK_{Q,P}) = {1\over Z_{Q,P}}  \int  \rd u \, {\rm e}^{ -u^2/2 \hat g_s}
\prod_{\alpha>0} 4 \sinh {u \cdot \alpha \over
2 P } \sinh {u \cdot \alpha \over
2 Q } {\rm ch}_R ({\rm e}^{u}).
\ee
In this equation, 
\be
\label{zpqpf}
Z_{Q,P}=\int  \rd u \, {\rm e}^{ -u^2/2 \hat g_s}
\prod_{\alpha>0} 4 \sinh {u \cdot \alpha \over
2 P } \sinh {u \cdot \alpha \over
2 Q },
\ee
the coupling constant $\hat g_s$ is
\begin{equation}
\hat g_s =PQ g_s, \qquad g_s={2 \pi \ri \over k+y},
\ee
$y$ is the dual Coxeter number of $G$, and $u$ is an element in $\Lambda_w \otimes \IR$. $\alpha>0$ are the positive roots. 
Notice that, although $Q,P$ are {\it a priori} integer numbers, the integral formula above makes sense for any $Q,P$. 

The easiest way to 
prove (\ref{tkmi}) is by direct calculation. In order to do that, we first calculate the integral 
\be
\label{fint}
\int  \rd u \, {\rm e}^{ -u^2/2 \tilde g_s}
\prod_{\alpha>0} 4 \sinh {u \cdot \alpha \over
2  } \sinh {u \cdot \alpha \over
2 f } {\rm ch}_R ({\rm e}^{u}), \qquad \tilde g_s = f g_s,
\ee
where $f$ is arbitrary. We will also denote 
\be
Z_{1,f}=\int  \rd u \, {\rm e}^{ -u^2/2 \tilde g_s}
\prod_{\alpha>0} 4 \sinh {u \cdot \alpha \over
2  } \sinh {u \cdot \alpha \over
2 f }.
\ee
Let $\Lambda_R$ be the highest weight associated to the representation $R$. Weyl's denominator formula and Weyl's formula for the character give, 
\be
\ba
\prod_{\alpha>0} 2 \sinh {u \cdot \alpha \over
2  } &=\sum_{w\in \CW} \epsilon(w) \re^{w(\rho)\cdot u}, \\
{\rm ch}_R(\re^u)&={ \sum_{w \in \CW} \epsilon(w) \re^{w(\rho+\Lambda_R)\cdot u} \over \sum_{w \in \CW} \epsilon(w) \re^{w(\rho)\cdot u}}, 
\ea
\ee
and the integral (\ref{fint}) becomes a sum of Gaussians, 
\be
\sum_{w, w'\in \CW} \epsilon(w)  \epsilon(w') \int \rd u \, \exp\left\{ -{u^2\over 2 \tilde g_s} + w(\rho+\Lambda_R)\cdot u + w'(\rho)\cdot u/f \right\}
\ee
Up to an overall factor which is independent of $\Lambda_R$ (and which will drop after normalizing by $Z_{1,f}$), this equals 
\be
\exp\left[ {g_s  f\over 2} \left( \Lambda_R+\rho\right)^2 \right] \sum_{w\in \CW} \epsilon(w)  \exp\left( g_s \rho\cdot w(\rho+ \Lambda_R)\right).
\ee
We then obtain
\be
\label{basicev}
\ba
& {1\over Z_{1,f}}\int  \rd u \, {\rm e}^{ -u^2/2 \hat g_s}
\prod_{\alpha>0} 4 \sinh {u \cdot \alpha \over
2  } \sinh {u \cdot \alpha \over
2 f } {\rm ch}_R ({\rm e}^{u})\\
& =
\exp\left[ {g_s  f\over 2}\left( \left(\Lambda_R+ \rho\right)^2-\rho^2\right) \right] {\sum_{w\in \CW} \epsilon(w)  \exp\left( g_s \rho\cdot w(\rho+ \Lambda_R)\right) 
\over \sum_{w\in \CW} \epsilon(w)  \exp\left( g_s \rho\cdot w(\rho)\right)}
=\re^{2 \pi \ri f h_R } {\rm dim}_q(R). 
\ea
\ee

With this result, it is trivial to evaluate (\ref{tkmi}). The change of variables $u= Q x$ leads to 
\be
\label{intint}
W_R(\CK_{Q,P}) = {1\over Z_{1,f}}  \int  \rd x \, {\rm e}^{ -x^2/2 \tilde g_s}
\prod_{\alpha>0} 4 \sinh {x \cdot \alpha \over
2  } \sinh {x \cdot \alpha \over
2 f } {\rm ch}_R ({\rm e}^{Q x}),
\ee
where 
\be
f=P/Q.
\ee
We can now expand ${\rm ch}_R ({\rm e}^{Q x})$ by using Adams' operation (\ref{adams}). The resulting sum can be evaluated by using (\ref{basicev}), and one obtains
\be
W_R(\CK_{Q,P}) = \sum_V c_{R,Q}^V \re^{2 \pi \ri Q/P h_V } {\rm dim}_q(V), 
\ee
which is exactly (\ref{finaltorus})\footnote{A direct calculation of the integral (\ref{tkmi}) is presented in \cite{tierztorus} by using the formalism 
of biorthogonal polynomials. The result seems to agree with the above calculations, but the framing factor is not clearly identified.}.  
Therefore, (\ref{tkmi}) is manifestly equal 
to the knot theory result, and in particular to the formula of Rosso and Jones for torus knots invariants. Notice that this matrix integral representation also comes with the 
natural framing $QP$ for the $(Q,P)$ torus knot. A similar calculation for $Z_{Q,P}$ shows that, up to an overall framing factor of the form 
\be
\exp\left[ {g_s \over 2} \left( {P \over Q} + {Q\over P}\right) \rho^2\right],
\ee
the partition function (\ref{zpqpf}) is independent of $Q,P$. This can be also deduced from the calculation in \cite{dt}.

We also note that there is an obvious generalization of the matrix model representation (\ref{tkmi}) to the torus link $(Q,P)$, given by 
\be
\label{torusint}
 W_{(R_1, \cdots, R_L)} (\CL_{Q,P})  = {1 \over Z_{Q/L,P/L}}
\int \rd  u  \, {\rm e}^{-u^2 /2 \hat g_s } \prod_{\alpha>0} 4 \sinh {u \cdot \alpha \over
2 P/L } \sinh {u \cdot \alpha \over
2 Q/L }  \prod_{j=1}^L {\rm ch}_{R_j} ({\rm e}^{u_i}).
\end{equation}

Since (\ref{tkmi}) can be calculated exactly at finite $N$, and the result is identical to (\ref{finaltorus}), what is the main interest of such a matrix model representation? As in the case of the Chern--Simons partition function on $\IS^3$, it makes possible to 
extract a geometric, large $N$ limit of the torus knot correlation functions, as we will now see. The fact that $Z_{Q,P}$ is independent of $Q,P$ up to a framing factor strongly suggests 
that the spectral curves for different $Q,P$ should be symplectic transforms of each other. We will verify this and derive in this way the results proposed in section 3. 

\subsection{Saddle--point equation}

We will now solve the matrix model (\ref{tkmi}), for the gauge group $U(N)$, and at large $N$. The first step is to derive the saddle--point equations 
governing the planar limit. 
An alternative route, which provides of course much more information, 
is to write full loop equations of the matrix model and then specialize them
to the planar part. This is presented in the Appendix. \\

As in \cite{tierz,tierztorus}, we first perform the change of variables
\be
u_i=PQ \log \, x_i, 
\label{eq:varlog}
\ee
which leads to 
\be
\ba
& \prod_{i=1}^N \rd u_i {\rm e}^{-\sum_i u^2_i/2 \hat g_s}\prod_{i<j}
\left( 2 \sinh {u_i - u_j\over
2 P} \right)\left( 2 \sinh {u_i - u_j\over
2 Q} \right)\\
&  =  (PQ)^N \prod_{i=1}^N \rd x_i \, \prod_{i<j} \left(x_i^Q -x_j^{Q}\right)\left(x_i^P -x_j^{P}\right)
\\ & \qquad \times  \prod_{i=1}^N\exp\left[- {PQ \over 2 g_s}  (\log x_i)^2 -\left( {P+Q\over 2}(N-1) +1\right) \log x_i \right] .
\ea
\ee
The matrix integral can then be written as
\be\label{matrixintegral1}
\ba
Z=(PQ)^N \int_{{\mathbb R_+^N}}\,\,\,\prod_{i=1}^N \rd x_i \, & \prod_{i<j} \left(x_i^Q -x_j^{Q}\right)\left(x_i^P -x_j^{P}\right)
\\ &  \times  \prod_{i=1}^N\exp\left[- {PQ \over 2 g_s}  (\log x_i)^2 -\left( {P+Q\over 2}(N-1) +1\right) \log x_i \right] ,
 \ea
\ee
where the variables $x_i={\rm e}^{u_i/PQ}$ are thought of as the eigenvalues
of a hermitian matrix $M$ of size $N\times N$, with only real positive
eigenvalues ($x_i\in \mathbb R_+$). \\

Define now  the resolvent
\be
\label{gresolvent}
G(x) = \tr \frac{x}{x-M} =  \sum_{i=1}^N {x\over x-x_i}.
\ee
Our observables are expectation values of product of resolvents, and their expansion into powers of $g_s$.
The 1-point function is
\be
W(x) = \left\langle G(x)\right\rangle = \sum_{g=0}^\infty g_s^{2g-1}\,W_g(x)
\ee
and its leading term $W_0(x)$ is called the spectral curve of the matrix model. The 2-point function is
\be
W_2(x_1,x_2) = \left \langle G(x_1)\,G(x_2)\right\rangle^{(c)}=\left\langle G(x_1)\,G(x_2)\right\rangle-\left\langle G(x_1)\right\rangle\,\left\langle G(x_2)\right\rangle = \sum_{g=0}^\infty g_s^{2g}\,W_g(x_1,x_2)
\ee
and similarly, the connected $n$-point correlation function is the cumulant of the expectation value of the product of $n$ resolvents
\be
W_n(x_1,\dots,x_n) = \left<G(x_1)\,G(x_2)\,\dots\,G(x_n)\ket^{(c)} = \sum_{g=0}^\infty g_s^{2g+n-2}\,W_g(x_1,\dots,x_n). 
\ee
We will denote in the following
\be
\omega = \exp\left( {2\pi \ri \over PQ} \right)
\ee
and the 't Hooft parameter of the matrix model is, as usual, 
\be
t=g_s N. 
\ee
\\

The saddle-point equations for the matrix integral (\ref{matrixintegral1}) are simply
\be
\label{EOMp}
\sum_{ j \not= i} \left[ {P x_i^{P} \over x_i^P -x_j^P} +  {Q x_i^{Q} \over x_i^Q -x_j^Q}\right]={PQ \over g_s} \log x_i +{P+Q\over 2}(N-1) +1.
\ee
If we use the identity
\be
{P x^{P-1} \over x^P -y^P}=\sum_{k=0}^{P-1} {1\over x-\omega^{-kQ} y},
\ee
we can write the first term in (\ref{EOMp}) as 
\be
\sum_{j\not=i} {2 x_i \over x_i -x_j} +  \sum_{k=1}^{P-1}\sum_{j\not=i} {x_i \over x_i-\omega^{-kQ} x_j} + \sum_{k=1}^{Q-1}\sum_{j\not=i} {x_i \over x_i-\omega^{-kP} x_j}.
\ee
But
\be
\sum_{j\not=i} {x_i \over x_i-\omega^{-kQ} x_j}  =\sum_{j=1}^N {x_i \over x_i -\omega^{-kQ} x_j}-{1\over 1-\omega^{-kQ}}= \frac{1}{g_s}W_0(x_i \omega^{k Q}) -{1\over 1-\omega^{-kQ}}, 
\ee
and the equation of motion reads, at leading order in $1/N$,
\be
\label{EOM}
PQ   \log x +{P+Q\over 2}\,t =W_0(x+\ri 0) + W_0(x-\ri 0) +\sum_{k=1}^{P-1} W_0(x\omega^{k Q}) + \sum_{k=1}^{Q-1} W_0(x\omega^{k P}).
\ee
This is the equation we will now solve. 

\subsection{Solving the saddle--point equations}
The resolvent $W_0(x)$ is analytic in $\mathbb C\setminus \CC$, where $\CC$ is a finite set of cuts in the complex plane. It satisfies
\be
\mathop{{\rm lim}}_{x\rightarrow 0} W_0(x) = 0, 
\ee
and 
\be
\mathop{{\rm lim}}_{x\rightarrow \infty}W_0(x) = t.
\ee
We now write the exponentiated version of the resolvent as
\be 
y=C_0\, x\, \re^{-{P+Q\over PQ}\, W_0(x)}
\ee
where 
\be
C_0= -\,\re^{{P+Q \over 2PQ}\,t}. 
\ee
$y$ is analytic in $\mathbb C\setminus \CC$ and satisfies the equation, 
\be
\label{disc}
y(x+\ri 0)\prod_{k=1}^{P-1} y(x \omega^{k Q}) \prod_{k=1}^{Q-1} y(x \omega^{k P}) ={1\over y(x-\ri 0)},
\ee
as well as the boundary conditions
\be
y(x) \sim C_0 x, \qquad x\rightarrow 0, 
\label{eq:aszero}
\ee
and
\be
y(x) \sim C_0^{-1} x, 
%\exp \left( -{P+Q \over PQ} t\right), 
\qquad x \rightarrow
\infty.
\label{eq:asinf}
\ee
Notice that $y$ vanishes only at $x=0$ and diverges only at $x=\infty$.

We now introduce the $P+Q$ functions 
\be
\ba
F_k(x)&=\prod_{l=0}^{P-1} y(x\,\, \omega^{kP+lQ}),  \qquad\quad 0\leq k\leq Q-1, \\
F_{Q+l}(x)&=\,\prod_{k=0}^{Q-1} \frac{1}{y(x\,\, \omega^{kP+lQ})}, \qquad\quad 0\leq l\leq P-1.
\ea
\label{eq:F}
\ee
If we assume that $y(x)$ has a single-cut $\CC$ on an interval $[a,b]$, then
$F_k$ has cuts  through the rotations of this cut by angles which are integer
multiples of $2\pi/PQ$. \\

If $0\leq k\leq Q-1$ and $0\leq l\leq P-1$, both $F_k$ and $F_{Q+l}$ have a cut across $\omega^{kP+lQ}\CC$, and according to (\ref{disc}), under crossing the cut we have
\be
F_k(x- \ri 0) = F_{Q+l}(x+ \ri0).
\ee
This implies that the function
\be
\CS(x,f) = \prod_{k=0}^{P+Q-1} (f-F_k(x))
\ee
has no cut at all in the complex plane:
\be
\CS(x+\ri0,f)=\CS(x-\ri0,f).
\ee
Its only singularities may occur when $y=\infty$ or when $y=0$ (indeed $y$ appears in the denominator in (\ref{eq:F})), and thus the only singularities are poles at $x=0$ or $x=\infty$. If we write 
\be
\CS(x,f)=\sum_{k=0}^{P+Q} (-1)^k\,\,\CS_k(x)\,f^{P+Q-k}
\ee
then each $\CS_k(x)$ is a Laurent polynomial of $x$. Besides, it is clear that 
\be
\CS (\omega x,f)=\CS(x,f),
\ee
therefore each $\CS_k(x)$ is in fact a Laurent polynomial in the variable $x^{PQ}$.
We clearly have
\be
\CS_0(x) =1
\,\, , \qquad
\CS_{P+Q}(x) = \prod_{k=0}^{P+Q-1} F_k(x)=1,
\ee
as well as the boundary conditions
\be\label{asympbehavFa}
0\leq k\leq Q-1\qquad \quad \begin{array}{l}
F_k(x) \sim -\,\, (-1)^P\,\, \omega^{kP^2}\,\, C_0^P x^P\,\,, \qquad x\rightarrow 0, \cr
F_k(x) \sim -\,\, (-1)^P\,\, \omega^{kP^2}\,\, C_0^{-P} x^P\,\,, \qquad x\rightarrow \infty, 
\end{array}
\ee
\be\label{asympbehavFb}
0\leq l\leq P-1\qquad \quad \begin{array}{l}
F_{Q+l}(x) \sim -\,(-1)^Q\,\, \omega^{-lQ^2}\,\, C_0^{-Q} x^{-Q}\,\,, \qquad x\rightarrow 0, \cr
F_{Q+l}(x) \sim  -\,(-1)^Q\,\, \omega^{-lQ^2}\,\, C_0^Q x^{-Q}\,\,, \qquad x\rightarrow \infty. 
\end{array}
\ee
This shows that the symmetric functions $\CS_k$ of the $F_k$'s must satisfy
\be
x\rightarrow 0
\qquad \quad
\left\{\begin{array}{ll}
1\leq k\leq P-1,\qquad & \quad \CS_k(x) =  \CO(x^{-kQ}), \cr
k=P,\qquad & \quad \CS_k(x) = (-1)^{P+Q}\,\, C_0^{-PQ}\,\, x^{-PQ} \,\, \left(1+\CO(x)\right),\cr
P+1\leq k\leq P+Q-1,\qquad & \quad \CS_k(x) =  \CO\left(x^{-PQ}x^{(k-P)P}\right),
\end{array}\right.
\ee
and
\be
x\rightarrow \infty
\qquad \quad
\left\{\begin{array}{ll}
1\leq k\leq Q-1,\qquad & \quad \CS_k(x) =  \CO(x^{kP}), \cr
k=Q,\qquad & \quad \CS_k(x) = (-1)^{P+Q}\,\, C_0^{-PQ}\,\, x^{PQ} \,\, \left(1+\CO(1/x)\right), \cr
Q+1\leq k\leq P+Q-1,\qquad & \quad \CS_k(x) =  \CO\left(x^{PQ}x^{-(k-Q)Q}\right).
\end{array}\right.
\ee
Since $\CS_k(x)$ are functions of $x^{PQ}$, the above behavior implies the following form for $\CS(x,f)$:
\be
\CS(x,f) = 
f^{P+Q} 
+ 1
+ (-1)^{Q} C_0^{-PQ}  x^{-PQ} \,f^Q
+ (-1)^P C_0^{-PQ} x^{PQ}\,f^P
+ \sum_{k=1}^{P+Q-1} s_k f^k 
\ee
where $s_k$ are constants. The functions $f=F_k(x)$ and $f=F_{Q+l}(x)$ must all obey this algebraic relationship between $x^{PQ}$ and $f$:
\be
\CS (x,f)=0.
\ee
We still have to determine the coefficients $s_k$.

\subsection{Derivation of the spectral curve}

Our matrix model (\ref{tkmi}) can be regarded as a perturbation of a Gaussian
matrix integral, and thus the resolvent should have only one cut, i.e. the
spectral curve must be rational. This determines the coefficients $s_k$. \\

Saying that an algebraic equation $\tilde\CS(x^{PQ},f)=0$ is rational means that there exists a rational parametrization of the solution.
Since the equation is of degree $2$ in $x^{PQ}$, this means that, for each $f$, we have two possible values for $x^{PQ}$, i.e. two points on the spectral curve. In other words, $f$ is a rational function of degree $2$ of an auxiliary parametric variable which we call $V$ (later we shall see that it indeed coincides with the function $V(x)$ defined in (\ref{tksc})). 
Upon a Moebius change of variable on $V$, we can always fix 3 points, and assume that $f$ has a pole at $V=\infty$ and at $V=1$, and a zero at $V=0$, i.e. we write it
\be
f = A\,\,V\,\,\frac{1-c^{-2}\,V}{1-V},
\ee
where the location of the second zero $c^2$ is to be determined later, but will eventually agree with the value given by the definition (\ref{cdef}).
Since the equation is of degree $P+Q$ in $f$, this means that, for each $x^{PQ}$, we have $P+Q$ values for $f$, i.e. $P+Q$ points on the spectral curve. We conclude that $x^{PQ}$ is a rational function of degree $P+Q$ of the auxiliary parametric variable $V$
\be
x^{PQ} = \CR_{P+Q}(V),
\ee
where $\CR_{P+Q}$ is a rational function with $P+Q$ poles. 
Moreover, the behavior at $x\to 0$ (i.e. at $V\to 0$ or $V\to c^2$) can be of the form
\be
f= F_k= \CO(x^P)=\CO(x^{PQ\over Q}).
\ee
Since $f$ is a rational function, it cannot behave like a fractional power, therefore $x^{PQ}$ must have a zero of an order which is a multiple of $Q$, let us say at $V=c^2$. 
The behavior of 
\be
f=F_{Q+l}=\CO(x^{-Q})
\ee
implies that $x^{PQ}$ must have a pole of an order which is a multiple of $P$, let us say at $V=0$.
Similarly, the behaviors at $x\to \infty$, i.e. $V\to 1$ or $V\to \infty$, imply that $x^{PQ}$ has a pole of an order which is a multiple of $Q$ and a zero of an order which is a multiple of $P$. Since the total degree of $x^{PQ}$ is $P+Q$, this means that  the orders of the poles and zeroes must be exactly $P$ and $Q$, respectively, and there are no other possible poles and zeroes.
We have then obtained that
\be
x^{PQ} = B\,\,V^{-P}\,\,\left(1-c^{-2}V\over 1- V\right)^Q.
\ee
Matching the behaviors of (\ref{asympbehavFa}) and (\ref{asympbehavFb}) gives the values of the coefficients $A,c, B$: 
\be
A=-1, \qquad c=\re^{t/2}, \qquad  B=c^{P+Q}, 
\ee
and in particular identifies $c=\re^{t/2}$ with the variable introduced in (\ref{cdef}). We finally obtain, 
\be\label{eqspcurve1}
\ba
x^{PQ} &= c^{P+Q} V^{-P}\,\,\left(1-c^{-2}V\over 1-\,V\right)^Q,\\
f &= -\,\,V\,\,{1-c^{-2}\,V\over 1-\,V}.
\ea
\ee
Notice that the relation between $X=x^{-PQ}$ and $V$ is precisely (\ref{polcurve}). \\

To complete the derivation of our spectral curve, let us recall the relationship between $f$ and the 
resolvent. 
We have, by definition of the resolvent,
\be
W(x) = \bra G(x) \ket = \bra \tr \frac{x}{x-M}\ket = \sum_{k=0}^\infty
x^{-k}\,\, \left<\tr M^k \ket= \sum_{k=0}^\infty x^{-k}\,\, \bra
\sum_{i=1}^N \re^{ku_i/PQ} \ket
\ee
and
\be
\ba
\sum_{l=0}^{P-1} W(\omega^{lQ}x)& = \sum_{l=0}^{P-1} \bra \tr \frac{x}{x-\omega^{-lQ}M}\ket = P\,\bra \tr \frac{x^P}{x^P-M^P} \ket\\
&= P\,\sum_{k=0}^\infty x^{-kP}\,\, \bra\sum_{i=1}^N \re^{ku_i/Q}\ket.
\ea
\ee
In the planar limit we obtain, 
\be
\ba
\bra \tr \frac{x^P}{x^P-M^P} \ket_{g=0} 
=& \frac{1}{P}\,\, \sum_{l=0}^{P-1} W_0(\omega^{lQ}x) \cr
=& -\,\frac{1}{P}\,\, \sum_{l=0}^{P-1} \frac{PQ}{P+Q}\,\ln{\frac{y(\omega^{lQ}x)}{C_0 \, \omega^{lQ}\,x}} \cr
=& -\,\frac{Q}{P+Q}\,\, \,\ln{\frac{F_0(x)}{(-1)^{P-1}\,C_0^P\,x^P}} \cr
=& -\,\frac{1}{P+Q}\,\, \,\ln{\frac{(-F_0(x))^Q}{(-C_0)^{PQ}\,x^{PQ}}}. \cr
\ea
\ee
In other words, the resolvent of $M^P$ is (up to trivial terms), the log of $F_0(x)$, which is one branch of the algebraic function $f$.
From our explicit solution (\ref{eqspcurve1}) we find that
\be
{(-f)^Q\over (-C_0)^{PQ}\,\,x^{PQ}} = (V\,\re^{-t})^{P+Q}
\ee
therefore
\be
\ba
\sum_{k=0}^\infty x^{-kP}\,\, \left\langle \sum_{i=1}^N \re^{ku_i/Q}\right\rangle = \bra \tr \frac{x^P}{x^P-M^P} \ket_{g=0} 
=& \,\,t\,- \,\ln{{V(x)}} .\cr
\ea
\ee
We have then proved that $t-\ln V(x)$ is the resolvent of $M^P$. It allows to compute expectation values of traces of powers of $M$ which are multiples of $P$. 
Our derivation also shows that the $X$ in (\ref{eq:defWnc}) is
\be
X= x^{-PQ}.
\ee
Since the relation between $X=x^{-PQ}$ and $V$ is precisely (\ref{polcurve}), we have derived the torus knot spectral curve from the matrix model. \\

We can also deduce from this derivation 
an interpretation for the coefficients of the fractional powers of $X$ appearing in the calculations of section 3: they compute the correlators of the more general operators 
\be
{\rm ch}_R\left(\re^{n u/Q} \right)
\ee
in the matrix model, which should correspond to ``fractional holonomies" around torus knots in Chern--Simons theory. Finally, we should mention that the same method used to derive the spectral curve makes possible in principe to compute the 2-point function (\ref{eq:w02}), and to prove the topological recursion (\ref{eq:02berg}).

\sectiono{Conclusions and prospects for future work}

In this paper we have proposed and derived a spectral curve describing torus knots and links in the B-model. The curve turns out to be a natural generalization 
of \cite{akv}: one has just to consider the full ${\rm Sl}(2,\IZ)$ group acting on the standard curve describing the 
resolved conifold. Our result fits in very nicely with the construction of torus knot operators in 
Chern--Simons gauge theory, and with the matrix model representation of their quantum invariants. 

As we mentioned in the introduction, the ingredients we use to deal with torus knots are the same ones that were used to deal with the framed unknot. 
Going beyond torus knots in the context of topological string theory (for example, the figure-eight knot) will probably require qualitatively new 
ingredients in the construction, but this is already the case in Chern--Simons 
gauge theory, where the colored invariants of generic knots involve the quantum $6j$ coefficients \cite{witten6j,rama}. We hope that the results for torus knots obtained in this paper 
will be useful to address the more general case.  

The structure we have found for the invariants of torus knots should have an A-model counterpart. In the A-model, framing arises as an ambiguity associated 
to the choice of localization action in the moduli space of maps with boundaries \cite{kl}, and the open string invariants depend naturally on an integer parametrizing this ambiguity. Our analysis indicates that there should be a two-parameter family of open string invariants generalizing the computations made for the unknot. 
These open string invariants can be in principle computed in terms of intersection theory on the moduli space of Riemann surfaces. As an example of this, consider 
the coefficient of the highest power of $c$ (which is $P+Q$) in the HOMFLY invariant of the $(Q,P)$ torus knot (\ref{wfund}) with framing $QP$. This coefficient, which we call $a_{P+Q}(q)$, can be expanded in a power series in $g_s$:
\be
a_{P+Q}(q)=\sum_{g \ge 0} a_{P+Q}^{(g)} g_s^{2g-1}. 
\ee
It is easy to see from the results of \cite{kl,mv} that the coefficients $a_{P+Q}^{(g)}$ are given by:
\be
a_{P+Q}^{(g)}={ (-1)^{g} Q  \over (Q-1)!} 
 \prod_{j=1}^{Q-1} 
(j+P)  \,  {\rm Res}_{u=0} 
\int_{{\overline M}_{g,1}} 
{c_g (\IE ^{\vee}(u))c_g(\IE ^{\vee} ((-P/Q-1)u)) c_g (\IE ^{\vee} (P u/Q))
\over u^2  (u- Q\psi_1)}.
\ee
In this formula, as in \cite{kl,mv}, ${\overline{M}}_{g,1}$ is the
Deligne--Mumford compactification of the moduli space of genus $g$, 1--pointed
Riemann surfaces, 
$\psi_1$ is the Chern class of the tautological line bundle $\mathbb{L}_1\to \overline{M}_{g,1}$, $\IE$ is the Hodge bundle over $\overline M_{g,1}$, and we denote 
\be
\label{seriesho}
c_g(\IE ^{\vee} (u))= \sum_{i=0}^g c_{g-i} (\IE ^{\vee}) u^i.
\ee
Although we have written down an example with $h=1$, the generalization to higher $h$ invariants, in the spirit of \cite{mv}, 
is immediate. Perhaps these formulae can lead to an explicit A-model description of torus knot invariants, and in particular shed some 
light on the proposals for the corresponding Lagrangian submanifolds made in \cite{lmv,taubes,koshkin}. Notice that, according to our description, 
the A-model invariants should involve some sort of fractional framing. Such
framings, in the context of A-model localization, have been considered in
\cite{df, Brini:2010sw, Brini:2011ij}. Of course, 
it should be possible to implement the general symplectic transformation we are considering directly in the topological vertex. 

Although in this paper we have focused on the spectral curve of the resolved
conifold, one can consider general ${\rm Sl}(2,\IZ)$ transformations of open string amplitudes defined by arbitrary spectral curves. In some cases, these transformations have a knot theory interpretation. For example, the outer brane in local $\IP^1\times \IP^1$ describes the unknot in $L(2,1)=\IR \IP^3$ \cite{bkmp}, and its general modular transformations should describe torus knots in this manifold. It would be also interesting to see what is the relation between the approach to torus knots in this paper and the recent work based on Hilbert schemes of singularities in $\IC^2$ \cite{os}. 

The topological recursion of \cite{eo}, which computes open and closed topological string amplitudes in toric Calabi--Yau manifolds, might have a generalization 
which gives the mirror of the refined topological vertex \cite{ikv} (recent work in this direction can be found in \cite{chengetal}). 
If the only data entering in this generalization 
turn out to be the same ones appearing in the original recursion (i.e., if the refinement only requires the knowledge of the spectral curve and of 
the natural differential on it, as it happens for example in the $\beta$ deformation \cite{beta}), 
then one should be able to use our spectral curve (\ref{tksc}) 
to refine the colored HOMFLY polynomial of torus knots. The resulting refinement should provide interesting 
information on the Khovanov homology of torus knots and might lead to a computation of their ``superpolynomial" \cite{gsv,dgr}, 
as well as of its generalizations to higher representations. 

Finally, the techniques developed in this paper to analyze the matrix model for torus knots will probably be very useful in order to understand the large $N$ limit of 
the more general matrix models for Seifert spheres introduced in \cite{mm}. 
Such a large $N$ limit would give a way to derive the dual string geometries to Chern--Simons theory in more 
general rational homology spheres --a dual which has remained elusive so far. 

\section*{Acknowledgements}
We would like to thank Gaetan Borot, Stavros Garoufalidis, Albrecht Klemm, S\'ebastien Stevan and Pawel Traczyk for useful discussions
and correspondence. The work of A. B. and M. M. is supported in part by FNS. 
The work of B. E.  is partly supported by the ANR project GranMa ``Grandes Matrices Al\'{e}atoires'' ANR-08-BLAN-0311-01, by the European Science Foundation through the Misgam program, by the Quebec government with the FQRNT, and the CERN.

\appendix
\section{Loop equations}
In this Appendix we derive the loop equations for the matrix integral (\ref{matrixintegral1}). In the following it is useful to notice that the resolvent 
$G(x)$ defined in (\ref{gresolvent}) satisfies
\be
\sum_{a=0}^{Q-1} G(\omega^{aP+bQ}x) 
= \sum_i {Q\,(\omega^{bQ}x)^Q\over (\omega^{bQ}x)^Q-x_i^Q}\,
\ee

As it is well-known, the method of loop equations consists in observing that an integral is unchanged under change of variables. 
In our case we shall perform the infinitesimal change of variable $x_i\to x_i+\epsilon \delta x_i +O(\epsilon^2)$ where
\be
\delta x_i = \sum_{a=0}^{Q-1}\sum_{b=0}^{P-1} {\omega^{aP+bQ}\,\, x\,x_i\over \omega^{aP+bQ}\,x-x_i}.
\ee
The loop equation, which computes the term of order 1 in $\epsilon$, can be written as
\be
\bra \delta \ln \Delta(x_i^Q)+\delta \ln \Delta(x_i^P)+\sum_i  {\partial \delta x_i\over \partial x_i}\ket
= \bra \sum_i V'(x_i)\,\delta x_i\ket
\ee
where 
\be
V'(x) = {PQ \over g_s} {\ln x \over x} + \left({P+Q \over 2}(N-1)+1\right){1\over x}.
\ee
We thus have to compute
\be
\ba
\delta \ln\Delta(x_i^Q)
=& Q\,\sum_{b=0}^{P-1}\sum_{i<j} (\omega^{bQ}x)^Q\, {{Q x_i^Q\over (\omega^{bQ}\,x)^Q-x_i^Q}-{Q x_j^Q\over (\omega^{bQ}\,x)^Q-x_j^Q}\over x_i^Q-x_j^Q}  \cr
=& Q^2\,\sum_{b=0}^{P-1}\sum_{i<j} (\omega^{bQ}x)^{2Q}\, {1\over (\omega^{bQ}\,x)^Q-x_i^Q}\,\,{ 1\over (\omega^{bQ}\,x)^Q-x_j^Q}  \cr
=& {1\over 2}\,\, \sum_{a=0}^{Q-1}\sum_{a'=0}^{Q-1}\sum_{b=0}^{P-1} G(\omega^{aP+bQ}x)\,G(\omega^{a'P+bQ}x) \cr
& - {Q^2\over 2}\,\sum_{b=0}^{P-1}\sum_{i} (\omega^{bQ}x)^{2Q}\, {1\over ((\omega^{bQ}\,x)^Q-x_i^Q)^2}.
\ea
\ee
We also have
\be
\ba
\sum_i  {\partial \delta x_i\over \partial x_i}
&= \sum_{b=0}^{P-1} \sum_{a=0}^{Q-1} \sum_i  {\partial \over \partial x_i}  \, \left( {(\omega^{aP+bQ}x)^2\over \omega^{aP+bQ} x-x_i} - \omega^{aP+bQ}x \right)  \cr
&= \sum_{b=0}^{P-1} \sum_{a=0}^{Q-1} \sum_i   \,  {(\omega^{aP+bQ}x)^2\over (\omega^{aP+bQ} x-x_i)^2}   \cr
&= -x^2\,{\partial \over \partial x}\,\sum_{b=0}^{P-1} \sum_{a=0}^{Q-1}  \sum_i \,  { \omega^{aP+bQ}\over (\omega^{aP+bQ} x-x_i)}   \cr
&= -x^2\,{\partial \over \partial x}\,\sum_{b=0}^{P-1}   \sum_i \,  { \omega^{bQ}\,\, Q\,(\omega^{bQ} x)^{Q-1} \over ((\omega^{bQ} x)^Q-x_i^Q)}   \cr
&= \sum_{b=0}^{P-1}   \sum_i \,  { \,\, Q^2\,(\omega^{bQ} x)^{2Q} \over ((\omega^{bQ} x)^Q-x_i^Q)^2}   
 -\sum_{b=0}^{P-1}   \sum_i \,  { \,\, Q(Q-1)\,( \omega^{bQ}\,x)^{Q} \over ((\omega^{bQ} x)^Q-x_i^Q)}   \cr
&= \sum_{b=0}^{P-1}   \sum_i \,  { \,\, Q^2\,(\omega^{bQ} x)^{2Q} \over ((\omega^{bQ} x)^Q-x_i^Q)^2}   
 - (Q-1)\,\sum_{a=0}^{Q-1}\sum_{b=0}^{P-1}  G(\omega^{aP+bQ}x) .\cr
\ea
\ee
The loop equation then gives
\be
\ba
& {1\over 2}\,\, \sum_{a=0}^{Q-1}\sum_{a'=0}^{Q-1}\sum_{b=0}^{P-1} \bra G(\omega^{aP+bQ}x)\,G(\omega^{a'P+bQ}x) \ket \cr
& + {1\over 2}\,\, \sum_{a=0}^{Q-1}\sum_{b=0}^{P-1}\sum_{b'=0}^{P-1} \bra G(\omega^{aP+bQ}x)\,G(\omega^{aP+b'Q}x) \ket \cr
&=  \sum_{a=0}^{Q-1}\sum_{b=0}^{P-1} \bra \sum_i \left( x_iV'(x_i)+{P+Q-2\over 2}\right)\,\,{\omega^{aP_+bQ} x\over \omega^{aP_+bQ} x-x_i} \ket,\cr
\ea
\ee
i.e.
\be
\ba
& {1\over 2}\,\, \sum_{a=0}^{Q-1}\sum_{a'=0}^{Q-1}\sum_{b=0}^{P-1} \bra G(\omega^{aP+bQ}x)\,G(\omega^{a'P+bQ}x) \ket 
 + {1\over 2}\,\, \sum_{a=0}^{Q-1}\sum_{b=0}^{P-1}\sum_{b'=0}^{P-1} \bra G(\omega^{aP+bQ}x)\,G(\omega^{aP+b'Q}x) \ket \cr
&=  \sum_{a=0}^{Q-1}\sum_{b=0}^{P-1} \bra \sum_i \left( {PQ \over g_s}  \ln{x_i} +{P+Q\over 2}N\right)\,\,{\omega^{aP_+bQ} x\over \omega^{aP_+bQ} x-x_i} \ket \cr
&= \sum_{a=0}^{Q-1}\sum_{b=0}^{P-1}\, \left( {PQ \over g_s} \ln{(\omega^{aP+bQ}x)} +{P+Q\over 2}N \right)  \bra G(\omega^{aP_+bQ} x) \ket \cr
& -  \sum_{a=0}^{Q-1}\sum_{b=0}^{P-1}  {PQ \over g_s} \,\omega^{aP_+bQ} x\,\bra \sum_i\,{\ln{(\omega^{aP+bQ}x)}-\ln{x_i}\over \omega^{aP_+bQ} x-x_i} \ket. \cr
\ea
\ee
This is then our main loop equation:
\be
\ba
& {1\over 2}\,\, \sum_{a=0}^{Q-1}\sum_{a'=0}^{Q-1}\sum_{b=0}^{P-1} \bra G(\omega^{aP+bQ}x)\,G(\omega^{a'P+bQ}x) \ket 
 + {1\over 2}\,\, \sum_{a=0}^{Q-1}\sum_{b=0}^{P-1}\sum_{b'=0}^{P-1} \bra G(\omega^{aP+bQ}x)\,G(\omega^{aP+b'Q}x) \ket \cr
&= \sum_{a=0}^{Q-1}\sum_{b=0}^{P-1}\, \left( {PQ \over g_s} \ln{(\omega^{aP+bQ}x)} +{P+Q\over 2}N \right)  \bra G(\omega^{aP_+bQ} x) \ket \cr
& -  \sum_{a=0}^{Q-1}\sum_{b=0}^{P-1}   {PQ \over g_s} \,\omega^{aP_+bQ} x\,\bra \sum_i\,{\ln{(\omega^{aP+bQ}x)}-\ln{x_i}\over \omega^{aP_+bQ} x-x_i} \ket \cr
\ea
\ee
We deduce that the spectral curve $W_0(x)$ satisfies
\be
\ba
& {1\over 2}\,\, \sum_{a=0}^{Q-1}\sum_{a'=0}^{Q-1}\sum_{b=0}^{P-1} W_0(\omega^{aP+bQ}x)\,W_0(\omega^{a'P+bQ}x)
 + {1\over 2}\,\, \sum_{a=0}^{Q-1}\sum_{b=0}^{P-1}\sum_{b'=0}^{P-1} W_0(\omega^{aP+bQ}x)\,W_0(\omega^{aP+b'Q}x)  \cr
&= g_s\sum_{a=0}^{Q-1}\sum_{b=0}^{P-1}\, \left( {PQ \over g_s}  \ln{(\omega^{aP+bQ}x)} +{P+Q\over 2}N\right)  W_0(\omega^{aP_+bQ} x)   -  P(x)
\ea
\ee
where
\be
P(x) = \sum_{a=0}^{Q-1}\sum_{b=0}^{P-1}   PQ \,\omega^{aP_+bQ} x\,\bra \sum_i\,{\ln{(\omega^{aP+bQ}x)}-\ln{x_i}\over \omega^{aP_+bQ} x-x_i} \ket. 
\ee

We will now assume that $W_0(x)$ has only one cut $\CC$ in the complex plane. $P(x)$ has no discontinuity through $\CC$, 
\be
P(x+\ri0)=P(x-\ri 0),
\ee
and the functions $W_0(\omega^{aP+bQ}x)$ with $(a,b)\neq (0,0)$ have no cut either, so it follows that
\be
\ba
W_0(x+\ri0)^2 + \sum_{a'=1}^{Q-1} W_0(x+i0) W_0(\omega^{a'P}x) + \sum_{b'=1}^{P-1} W_0(x+\ri0) W_0(\omega^{b'Q}x) \cr
- \, \left( PQ \ln{x} +{P+Q\over 2}t \right)  W_0(x+\ri0) \cr
= 
W_0(x-\ri0)^2 + \sum_{a'=1}^{Q-1} W_0(x-\ri0) W_0(\omega^{a'P}x) + \sum_{b'=1}^{P-1} W_0(x+\ri0) W_0(\omega^{b'Q}x)\cr
 - \, \left( PQ \ln{x} +{P+Q\over 2}t \right)  W_0(x-\ri0),
\ea
\ee
i.e., if we put all terms in the left hand side and divide by $W(x+\ri0)-W(x-\ri0)$, we get the equation (\ref{EOM}) which we derived with the saddle-point method.

\end{document}